\documentclass[a4paper,fleqn,usenatbib]{mnras}
\usepackage{newtxtext,newtxmath}
\usepackage[T1]{fontenc}
\usepackage{ae,aecompl}
\usepackage{graphicx}	% Including figure files
\usepackage{amsmath}	% Advanced maths commands
\usepackage{amssymb}	% Extra maths symbols

\title[NGC 1052: Gas excitation and kinematics]{A panchromatic spatially resolved study of the inner 500\,pc of NGC\,1052 - II: Gas excitation and kinematics}

\author[Dahmer-Hahn et al.]{L. G. Dahmer-Hahn $^{1}$\thanks{E-mail: dahmer.hahn@ufrgs.br}, R. Riffel$^1$, T.V. Ricci$^2$, J. E. Steiner$^3$, T. Storchi-Bergmann$^1$, \newauthor R. A. Riffel$^{4,5}$,  R. B. Menezes$^6$, N.Z. Dametto$^7$,  M. R. Diniz$^4$, J. C. Motter$^1$, \newauthor D. Ruschel-Dutra$^7$ 
\\
$^{1}$Departamento de Astronomia, Universidade Federal do Rio Grande do Sul. Av. Bento Goncalves 9500, 91501-970, Porto Alegre, RS, Brazil.\\
$^{2}$Universidade Federal da Fronteira Sul, Campus Cerro Largo, RS 97900-000, Brazil\\
$^{3}$Instituto de Astronomia, Geof\'isica e Ci\^encias Atmosf\'ericas, Universidade de S\~ao Paulo, 05508-900, S\~ao Paulo, Brazil\\
$^{4}$Universidade Federal de Santa Maria, Departamento de F\'isica, Centro de Ci\^encias Naturais e Exatas, 97105-900, Santa Maria, RS, Brazil\\
$^{5}$Center for Astrophysical Sciences, Department of Physics and Astronomy, The Johns Hopkins University, Baltimore, MD 21218, USA\\
$^{6}$Centro de Ci\^encias Naturais e Humanas, Universidade Federal do ABC, 09210-580, SP, Brazil\\
$^{7}$Departamento de F\'isica - CFM - Universidade Federal de Santa Catarina,  476, 88040-900, Florian\'opolis, SC, Brazil
}

\date{Accepted XXX. Received YYY; in original form ZZZ}

\pubyear{2019}

\begin{document}
\label{firstpage}
\pagerange{\pageref{firstpage}--\pageref{lastpage}}
\maketitle

\begin{abstract}

We map the optical and near-infrared (NIR) emission-line flux distributions and kinematics of the inner 320$\times$535\,pc$^2$ of the elliptical galaxy NGC\,1052. The integral field spectra were obtained with the Gemini Telescope using the GMOS-IFU and NIFS instruments, with angular resolutions of 0\farcs88 and 0\farcs1 in the optical and NIR, respectively. We detect five kinematic components: (1 and 2) Two spatially unresolved components, being a broad line region visible in H$\alpha$, with a FWHM of $\sim$3200\,km\,s$^{-1}$ and an intermediate-broad component seen in the [O{\sc iii}]$\lambda \lambda$4959,5007 doublet; (3) an extended intermediate-width component with 280<FWHM<450\,km\,s$^{-1}$ and centroid velocities up to 400\,km\,s$^{-1}$, which dominates the flux in our data, attributed either to a bipolar outflow related to the jets, rotation in an eccentric disc or a combination of a disc and large-scale gas bubbles; (4 and 5) two narrow (FWHM<150\,km\,s$^{-1}$) components, one visible in [O{\sc iii}], and one visible in the other emission lines, extending beyond the field-of-view of our data, which is attributed to large-scale shocks. Our results suggest that the ionization within the observed field of view cannot be explained by a single mechanism, with photoionization being the dominant mechanism in the nucleus with a combination of shocks and photoionization responsible for the extended ionization.

\end{abstract}


\section{Introduction}
\label{sec:introduction}

Depending on the main ionization source of a galaxy, different emission line intensities can be produced. While, for example, very young stellar populations tend to strongly ionize hydrogen if compared to other emission lines, active galactic nuclei (AGN) produce, among others, stronger optical [O{\sc iii}] and [N{\sc ii}] lines \citep[][hereafter K01 and K03]{kewley+01,kauffmann+03} if compared to hydrogen. Emission line ratios are, therefore, often used to discriminate between the possible ionization sources of galaxies \citep[e.g.][hereafter BPT]{bpt}.
\par
A special class of galaxies is the Low Ionization Nuclear Emission Line Region \citep[LINER,][]{heckman80}, which is characterized by high values of [N{\sc ii}]$\lambda$6583\r{A}/H$\alpha$, but lower values of [O{\sc iii}]$\lambda$5007\r{A}/H$\beta$ if compared to Seyfert galaxies and quasars (K03).
\par
The nature of the LINERs is still matter of intense debate \citep[e.g.][]{kehrig+12,papaderos+13,belfiore+15,gomes+16,belfiore+16}, since many mechanisms are able to produce a LINER-like emission line spectrum besides low luminosity AGN \citep{FerlandNetzer83,HalpernSteiner83}, such as shocks \citep[as initially proposed by][]{heckman80}, post-asymptotic giant branch stars (pAGB) \citep{binette+94, stasinska+08, CF11, YanBlanton12} and starbursts with ages between 3 and 5 Myr, dominated by Wolf-Rayet stars \citep{BarthShields00}.

\par

The hypothesis of a LINER powered by accretion onto a supermassive black hole has been confirmed in some sources. \citet{ho+97} showed that, for a sample of 211 emission-line nuclei, about 20\% have a broad component in their H$\alpha$ line, with most of these objects belonging to the LINER class.
\par
Recent studies using integral field unit (IFUs) data were able to isolate the components of the galaxies, searching for eventual central sources. From the SAURON sample, \citet{sarzi+10} found a tight correlation between the stellar surface brightness and the flux of the H$\beta$ recombination line. They also reported that hot evolved stars, probably pAGB stars, are the best candidates to ionize the gas. \citet{LoubserSoechting13} found, for a sample of four central cluster galaxies (CCG) with LINER emission, that AGN photoionization models (with higher metallicity) are able to reproduce their spatially resolved line ratios, although they could not rule out models with shocks or photoionization by pAGB stars. Also,  \citet{ricci+14a,ricci+14b,ricci+15} found in a sample composed of 10 LINERs, that most of them have emission compatible with a central AGN, although three of them have a circumnuclear structure with the shape of a disc, compatible with photoionization by pAGB stars and in one galaxy the ionization is compatible with shocks. By studying a sample of 14 Seyfert/LINER galaxies, \citet{belfiore+15} reported that the observed extended emission is also consistent with ionization from hot evolved stars. A similar scenario was found in a series of papers \citep{kehrig+12, papaderos+13, gomes+16}, where the authors used low spatial resolution data from Calar Alto Legacy Integral Field Area Survey, and found evidence of two different types of early-type galaxies (ETGs), which they classified as Type i and Type ii. A Type i ETG is a system with a nearly constant H$\alpha$ equivalent width [EW(H$\alpha$)] in their extranuclear component, compatible with the hypothesis of photoionization by pAGB as the main driver of extended warm interstellar medium (wim) emission, whereas  type ii ETGs are virtually wim-evacuated, with a very low outwardly increasing EW(H$\alpha$) ($\leq$ 0.5\r{A}).
\par
In order to further our understanding of galaxies with LINER emission, we conducted a study case for NGC\,1052, an E4 galaxy at a distance of 19.1$\pm$1.4\,Mpc, subject of a long and intense debate surrounding its LINER emission. For example, \citet{KoskiOsterbrock76} initially suggested that the gas ionization of the inner 2\farcs7$\times$4\farcs0 is due to shocks, with \citet{fosbury+78} presenting a detailed radiative shock model and \citet{fosbury+81} finding no evidence of a compact source of non-stellar radiation capable of ionizing enough gas to produce the observed Balmer-line fluxes. On the other hand, \citet{diaz+85} argued that the strengths of the [S{\sc iii}]$\lambda$9069,$\lambda$9532 lines favor photoionization. Later, \citet{SugaiMalkan00} studied NGC\,1052 using mid infrared spectra and found evidence supporting shocks as the origin of $\sim$80\% of the excitation of this source. \citet{gabel+00} also studied NGC~1052 and reported that emission-line fluxes obtained with the Faint Object Spectrograph attached to the Hubble Space Telescope can be simulated with simple photoionization models using a central source with a power law with spectral index $\alpha$ = -1.2. They also noted that simple model calculations using a gas with constant density do not match the intrinsic emission-line spectrum of NGC~1052 for any choice of density, requiring at least two different densities. Also, \citet[][ hereafter D15]{dopita+15} observed this galaxy with the IFS mode of the Wide Field Spectrograph, and reported the detection of two buoyant gas bubbles with $\sim$1.5\,kpc (15\farcs0) extending to both sides of the nucleus, which are expanding along the minor axis of the galaxy. They also found that, since two distinct densities can be found, a double-shock model explain better the data.
\par
Broad Hydrogen emission line components were also reported for NGC\,1052.  \citet{barth+99} used polarized light and confirmed a hidden Broad Line Region (BLR) on the H$\alpha$ line. Later, \citet[][hereafter S05]{sugai+05} observed this galaxy with the IFS mode of the Kyoto 3DII instrument mounted on the Subaru telescope, covering a $\sim$3\farcs0$\times$3\farcs0 FoV at $\sim$0\farcs4 spatial resolution. They reported direct detection of a broad H$\beta$ component, also finding evidence of three main kinematical components for the gas: a high-velocity bipolar outflow, low-velocity disc rotation, and a spatially unresolved nuclear component.
\par
In radio wavelengths, this galaxy displays two jets with slightly different orientations. On kiloparsec scales, \citet{wrobel84} found a radio jet oriented along the East-West direction, whereas on parsec scales, \citet{FeyCharlot97} found a radio jet slightly bent toward the North-South direction. Recently, \citet{baczko+16} calculated the parsec jet expansion velocity as $\beta$ = v/c = 0.46 $\pm$ 0.08 and $\beta$ = 0.69 $\pm$ 0.02 for the western and eastern jet, respectively.
\par
In X-ray wavelengths, according to \citet{kadler+04}, NGC 1052 displays a compact core, best fitted by an absorbed power law with column density $N_H$ = (0.6-0.8)$\times 10^{22}$ cm$^{-2}$. Besides, they also reported various jet-related emissions and an extended region, also aligned with the radio synchrotron jet-emission.
\par
In the first paper of this series \citep[][hereafter paper I]{luisgdh+19}, aimed at shedding some light on the understanding of the physical mechanism behind the gas excitation in NGC\,1052, we studied optical and near-infrared (NIR) stellar population properties of the inner 320$\times$535\,pc$^2$ of this galaxy. In Paper I, we fitted the stellar population simultaneously GMOS and NIFS datacubes for this galaxy. When using  only the optical data, we found that its central region is dominated by old (t$>$10\,Gyr) stellar populations, while when using NIR data we also found that the nucleus is dominated by an old stellar population but shows in addition a younger circumnuclear ring ($\sim$2.5\,Gyr). In the combined optical and NIR datacube we found  a dominance of older stellar populations. We also obtained stellar kinematics and we found that the stellar motions are dominated by a high ($\sim$240km\,s$^{-1}$) nuclear velocity dispersion, with stars also rotating around the center. Lastly, we measured equivalent widths of absorption features, both in the optical and in the NIR and found a drop in their values in the central regions of our FoV. We attributed this drop to the contribution of a featureless continuum emission from the low luminosity AGN (LLAGN).
\par
Here in the second paper, we map the physical properties of the emitting gas using optical and near-infrared data free from the star light contamination. The data containing only the gas emission were obtained after subtracting the contribution of the stellar populations to the observed fluxes, as derived in the first paper. We looked for possible ionization mechanisms to explain the emission lines within the central 3\farcs5$\times$5\farcs0 of NGC 1052. In order to achieve this, we use two main methods: direct emission-line fits, by identifying the kinematic components and obtaining spatially-resolved emission-line ratios, and the principal component analysis \citep[PCA,][]{pca} tomography technique. We also use literature radio data in order to search for correlations between the radio jet directions and our optical and NIR emission lines.
\par
This paper is structured as follows: in Section~\ref{sec:data}, we present the data and its reduction, while in Section~\ref{sec:fitting} we describe the emission line analysis. In Section~\ref{sec:results}, we present the results and in section~\ref{sec:discussion} we discuss them. Lastly, in Section~\ref{sec:conclusions}, we briefly summarize the main results found.

\section{Data And Reduction}
\label{sec:data}

\subsection{Optical Data}
The data acquisition and reduction are detailed in Paper I. In short, optical data were obtained using the Gemini Multi-Object Spectrograph (GMOS) on IFU mode. The original Field of View is 3\farcs5$\times$5\farcs0, with a natural seeing of 0\farcs88. The data were obtained using the B600 grating, resulting in a constant optical spectral resolution of $\sim$1.8\,\r{A}. The reduction process and data treatment were performed following the steps described in \citet{menezes+19}: trimming, bias subtraction, bad pixel and cosmic ray removal, extraction of the spectra, GCAL/twilight flat correction, wavelength calibration, sky subtraction, flux calibration, correction of the differential atmospheric refraction, high spatial-frequency components removal with the Butterworth spatial filtering, ``instrumental fingerprint'' removal and Richardson-Lucy deconvolution. The final angular resolution is 0\farcs70. The integrated optical continuum image, as obtained from the integration of the datacube, is shown in the left part of panel c of Fig.~\ref{fig:fig1}.
\subsection{NIR Data}
The Near-Infrared (NIR) data were obtained using Gemini North Near-Infrared Integral Field Spectrograph (hereafter NIFS) with the ALTAIR adaptive optics system. The angular resolution of the raw data is 0\farcs1$\times$0\farcs1. Since we applied a boxcar filter to improve the Signal-to-Noise ratio (S/N), the final spatial resolution is 0\farcs15$\times$0\farcs15. The spectral resolution is $\lambda/\Delta\lambda$ = 6040 for the J band and 5290 for the K band, corresponding to 50 and 57\,km\,s$^{-1}$ respectively. The data were reduced using the standard reduction scripts distributed by the Gemini team, which included trimming of the images, flat fielding, sky subtraction, wavelength and s-distortion calibrations and telluric absorption removal, flux calibration, differential atmospheric refraction correction, Butterworth spatial filtering and instrumental fingerprint removal. The final spectral coverage of the NIR data is 11472-13461\,\r{A} for the J band and 21060-24018\,\r{A} for the K band, and the final field of view is 2\farcs5$\times$2\farcs5. More details of the data reduction procedure can be found in Paper I. The NIR K-band continuum image, as obtained from the integration of the datacube, is shown in the right map of panel c of Fig.~\ref{fig:fig1}.
\subsection{Radio Data}
\begin{figure*}
	\includegraphics[width=\textwidth]{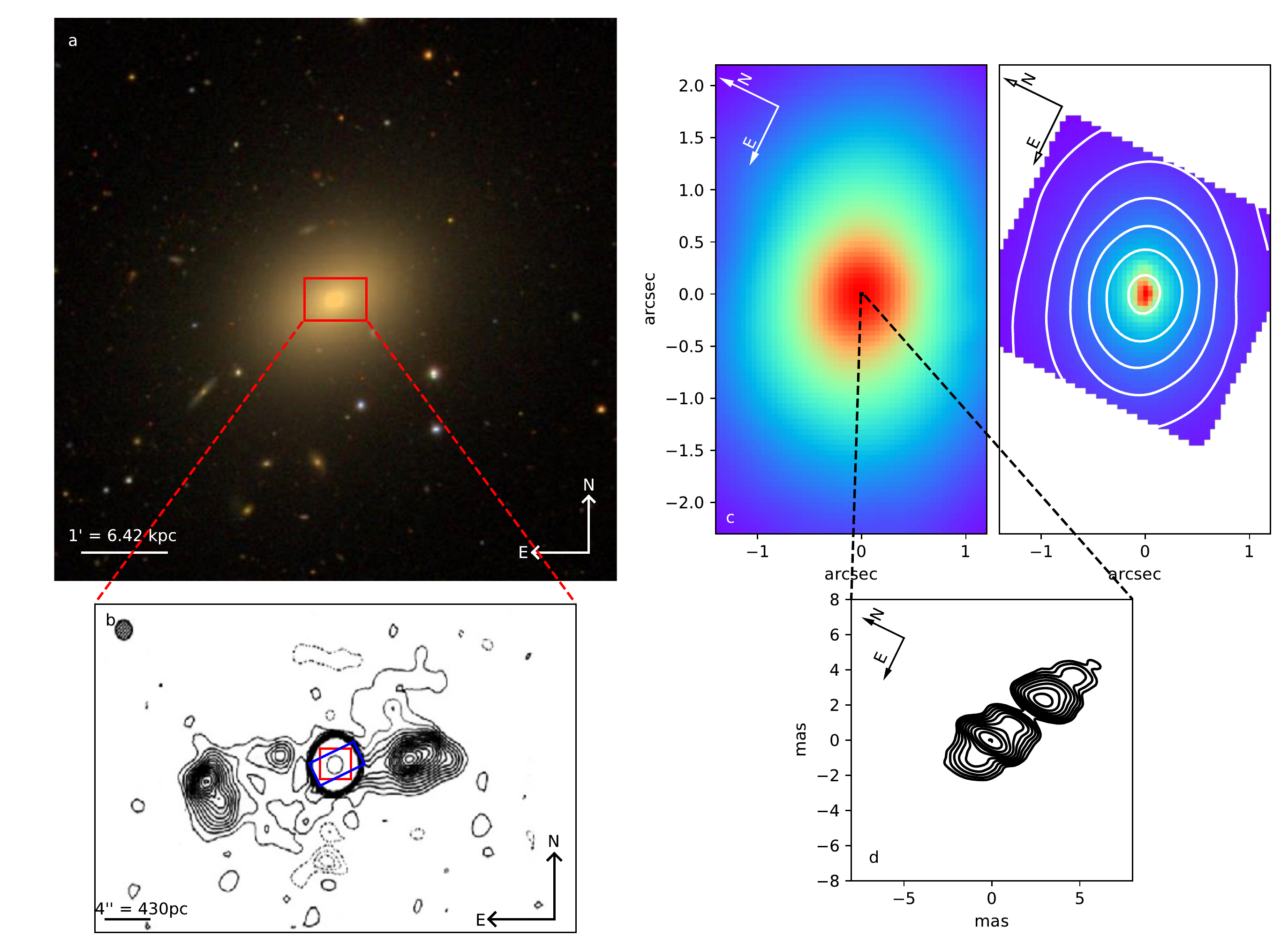}
	\caption{Panel a: SDSS image of NGC 1052 with the field of view of the panel b radio image in red. Panel b: NGC\,1052 total intensity map obtained with the VLA at 20\,cm \citep{wrobel84}. The map peaks at 737\,mJy/beam and contours are drawn at 500\,$\mu$Jy/beam x (-4,-3,-2,-1,1,2,3,4,5,6,7,8,9,10,11,12,13,14,15,800). Over this panel we indicate the GMOS FOV in blue and NIFS FOV in red. Panel c: left, optical continuum obtained with GMOS. In black, FOV of panel d radio image. Right, NIR k-band continuum obtained using NIFS. In white, optical contours of the GMOS-IFU continuum. Panel d: NGC 1052 total intensity map of the parsec scale jet obtained with the VLBA at 2\,cm on 2012/March/27 \citep{lister+18}. The bottom I contour is 1/512 of the peak of 0.551\,Jy/beam, and the contours increase in steps of 2. The restoring beam dimensions are 1.40$\times$0.50\,mas at the position angle -6$^{\circ}$.6.}
	\label{fig:fig1}
\end{figure*}

In order to assess any correlation between the gas kinematics probed by our optical and NIR observations with the direction of the NGC\,1052 radio jets, we show in Fig.~\ref{fig:fig1} (panel b) the historical 20 cm total intensity map of its kiloparsec scale jet obtained by \citet{wrobel84} using the Very Large Array (VLA). In Fig.~\ref{fig:fig1} (panel d), we show the 2\,cm total intensity map of the parsec scale jet of NGC\,1052 obtained using the Very Long Baseline Array (VLBA) and made publicly available by the MOJAVE team \citep[Monitoring of Jets in AGNs with VLBA Experiments,][]{lister+18}.

\section{Emission line fitting}
\label{sec:fitting}

The emission line fluxes were measured in a pure emission line spectrum, free from the underlying stellar flux contributions.  In order to obtain it, we used the optical and NIR stellar contents derived in Paper I, with E-MILES \citep{e-miles} models and the {\sc starlight} code \citep{CF04,CF05}. In Paper I, optical and NIR stellar populations were modeled separately, but were also derived after combining both ranges. We chose to subtract the stellar populations which were derived separately for optical and NIR datacubes, since NIR data was degraded in order to match the optical resolution. Additionally, in order to remove lower-order noise, we fitted and subtracted a five degree polynomial function to each spectrum after subtracting the stellar content. The systemic velocity was derived from the modelling of the stellar kinematics from Paper I.
\par
Close to the center of our FoV, a broad component is present in H$\alpha$. Since its distribution is unresolved by our data, we fitted this component separately, masking out the narrow components of [N{\sc ii}] and H$\alpha$, and forcing its spatial distribution to be the same of the PSF. The determined FWHM of this broad component is $\sim$3200\,km\,s$^{-1}$.
\par
After subtracting this broad component, the emission line fitting was performed with {\sc ifscube} \footnote{publicly available in the internet \url{https://bitbucket.org/danielrd6/ifscube}}, which is a Python based package of spectral analysis routines. The code allows the simultaneous fitting of multiple Gaussian or Gauss-Hermite profiles in velocity space, with or without constraints or bounds. The algorithm includes integrated support for pixel-by-pixel uncertainties, weights and flags, subtraction of stellar population spectra, pseudo-continuum fitting, signal-to-noise ratio evaluation and equivalent width measurements. The actual fitting relies on {\sc scipy}'s implementation of Sequential Least Squares Programming.
\par
Because of the complex kinematics of NGC\,1052, two Gaussian functions were needed in order to fit each emission-line profile, one narrow (100 < FWHM < 150\,km\,s$^{-1}$) and one with intermediate-width (hereafter IW, 280 < FWHM < 450\,km\,s$^{-1}$). We tried more complex configurations, such as using two gauss-hermite polynomials or three gaussians for each emission line, but none of them contributed significantly to the quality of the fits.
\par
Also, because of the complexity of the [N{\sc ii}] + H$\alpha$ region, where six gaussian functions were needed in a short wavelength range, we chose to fit the whole optical spectrum at the same time, constraining the FWHM and centroid velocity of emission lines with similar ionization potential, critical density, and which are formed close to each other in the nebulae. We divided the emission lines in four groups, constrained to have the same kinematics as follows:
\begin{enumerate}
    \item Ionized hydrogen (H$\alpha$, H$\beta$ and H$\gamma$)
    \item Doubly ionized oxygen ([O{\sc iii}] $\lambda$4363, $\lambda$4959 and $\lambda$5007\,\r{A})
    \item Neutral oxygen ([O{\sc i}] $\lambda$6300 and $\lambda$6360\,\r{A})
    \item Neutral and ionized nitrogen, and ionized sulfur ([N{\sc i}]$\lambda$5200\r{A}, [N{\sc ii}] $\lambda$5755,6548,6583\r{A} and [S{\sc ii}] $\lambda$6716,6731\r{A})
\end{enumerate}

We also performed the fitting using different configurations, such as forcing all emission lines to share the same double gaussian profile, or separating all emission lines. However, all these configurations point toward the same results, where only [O{\sc iii}] behaves differently compared to the other emission lines.

\section{Results}
\label{sec:results}
\subsection{Fluxes and Kinematics}

\begin{figure*}
	\includegraphics[width=\textwidth]{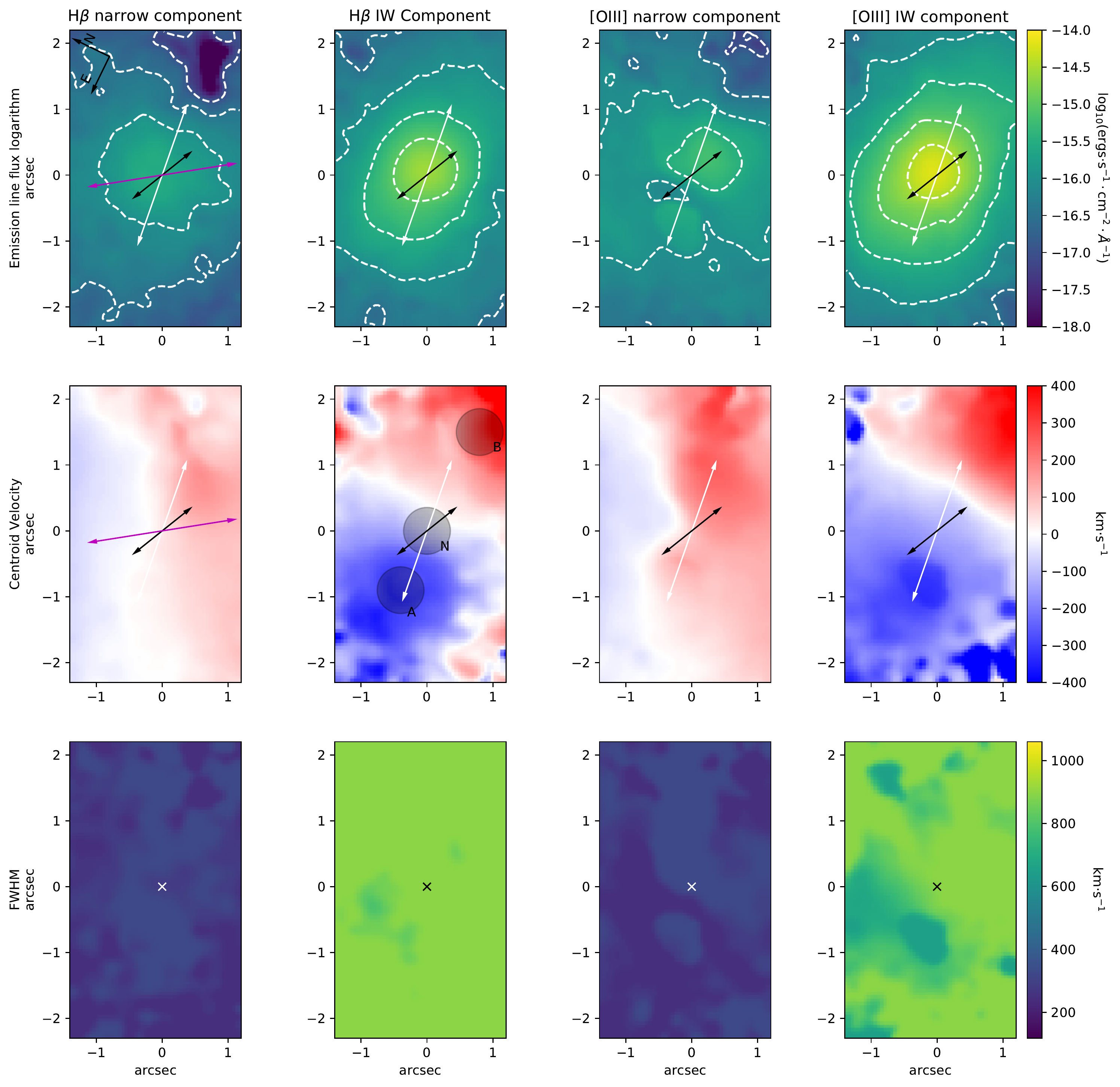}
	\caption{Emission line flux, centroid velocity and FWHM maps of H$\beta$ and [O{\sc iii}]$\lambda$5007\r{A} narrow and IW emission lines. The size of the spaxels is 0\farcs05. First row shows emission line flux maps in log scale with contours drawn. We superimpose on the H$\beta$ IW flux panel the three regions whose spectra are presented in Fig~\ref{fig:fits}. Middle row panels show centroid velocity maps, where rest-frame emission is white. In the middle row panels, the white and black arrows represent the orientations of the kiloparsec and parsec scale jet, respectively. In the narrow H$\beta$ centroid velocity map, the arrow in magenta represents the orientation of the gas bubbles identified by D15.}
	\label{fig:EL}
\end{figure*}

The flux distribution maps of the narrow and IW components of H$\beta$ and [O{\sc iii}]$\lambda$5007\r{A} emission lines are shown in the top panels of Fig~\ref{fig:EL}. We decided not to show the [O{\sc i}] , [N{\sc i}], [N{\sc ii}] and [S{\sc ii}] maps because they follow both the kinematics and flux distribution of the H{\sc i} lines. Also, we decided to plot H$\beta$ maps instead of H$\alpha$ since the later is blended with two [N{\sc ii}] emission lines. Over these panels, we show the orientation of the kiloparsec and parsec scale radio jets, represented by the white and black arrows, respectively. Over the narrow H$\beta$ centroid velocity map, we also show in magenta the orientation of the gas bubbles identified by D15.
\par
\begin{figure*}
	\includegraphics[width=\textwidth]{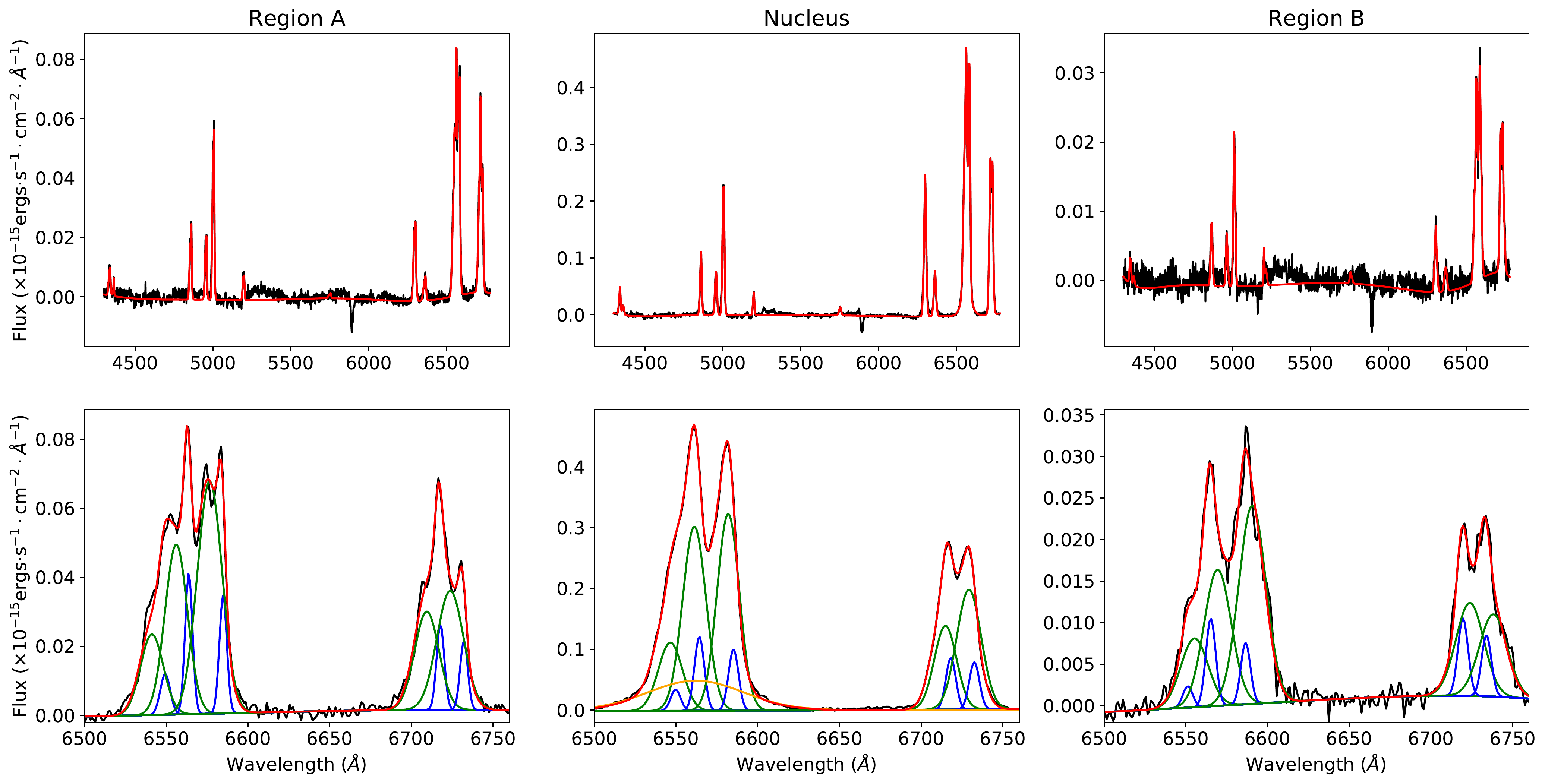}
	\caption{Fits of the optical emission lines of the central spaxel of the three regions marked in Fig.~\ref{fig:EL} after subtracting the contribution from the stellar population. The upper panels show the observed (black) and modeled spectrum (red). The bottom panels show in more detail the region between 6500 and 6760\,\r{A}. The different components of the modeled spectrum are shown in the bottom panels: narrow components are plotted in blue, IW components in green and the broad component is plotted in orange (central panel only).}
	\label{fig:fits}
\end{figure*}

%These maps are presented in log scale, with contours drawn. 
Although NGC\,1052 has complex gas kinematics throughout the entire FoV, we selected three key regions where this complexity is more evident. We labeled them N, for the nuclear region, and A and B, which lie in the direction of the kiloparsec scale jet. In the second column, first row panel of Fig.~\ref{fig:EL}, we indicate these three regions. The spectra and fits of the emission lines of these three regions are presented in Fig.~\ref{fig:fits}.
\par

In the middle and bottom panels of Fig~\ref{fig:EL}, we present the centroid velocity and FWHM maps of the narrow and IW components of H$\beta$ and [O{\sc iii}]$\lambda$5007\r{A} emission lines. We also overploted on the centroid velocity panels the orientations of kiloparsec and parsec scale radio jets in white and black arrows, and in magenta the orientation of D15 gas bubbles. The regions A, B and N are also marked over the H$\beta$ IW centroid velocity panel.
\par
\begin{figure}
	\includegraphics[width=\columnwidth]{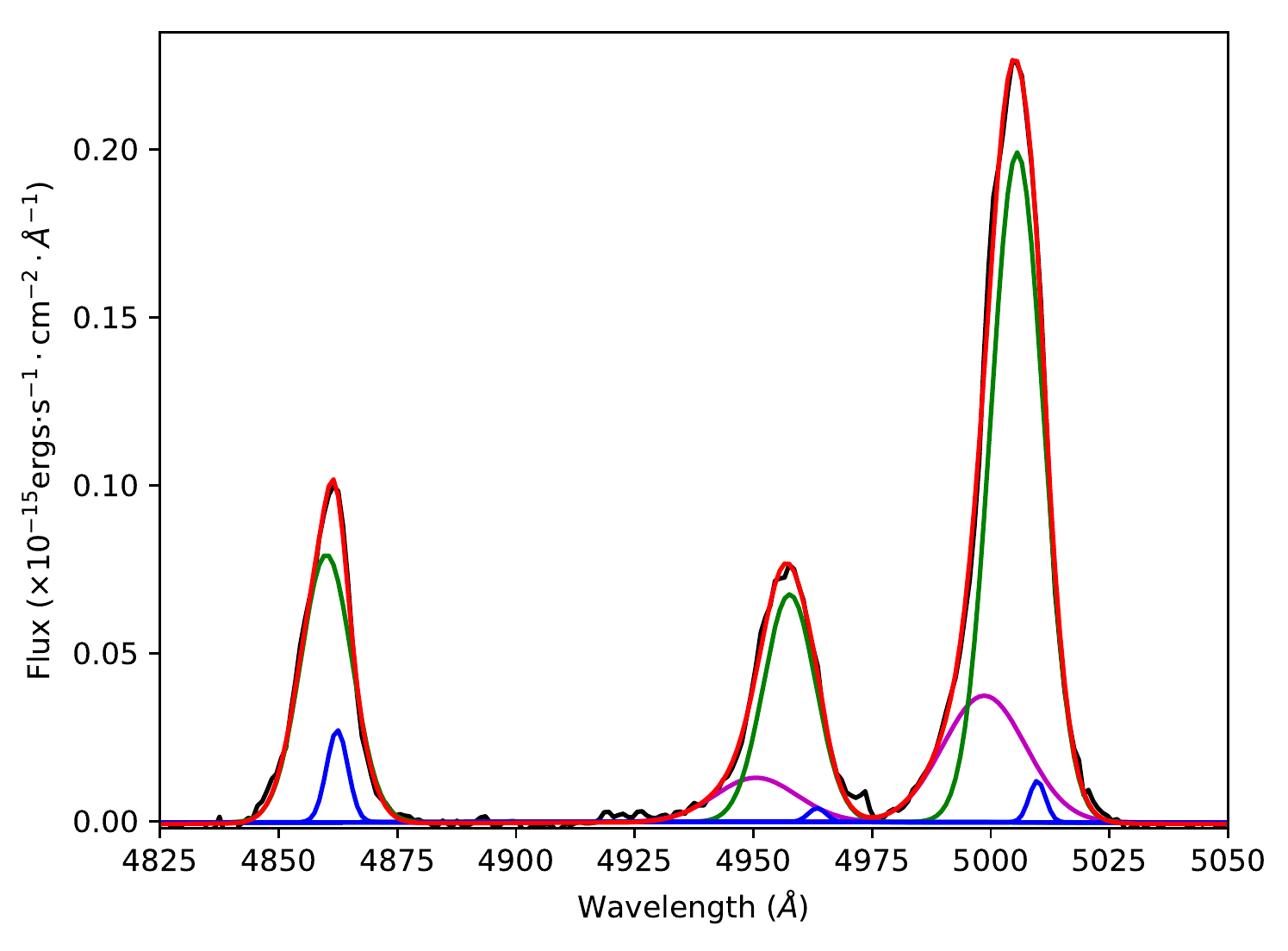}
	\caption{Fits of the H$\beta$ and nebular [O{\sc iii}] emissions of the central spaxel of region N, after subtracting the contribution from the stellar population. The  observed and modeled spectra are shown in black and red, respectively. Narrow components are plotted in blue, IW components in green, and the [O{\sc iii}] broad-intermediate components are plotted in magenta.}
	\label{fig:Oiii}
\end{figure}

In the nuclear region, a blue wing is visible in the nebular [O{\sc iii}] emission lines. A third gaussian function was needed in order to fit this region. We show in Fig.~\ref{fig:Oiii} the spectra and fits of the [O{\sc iii}] and H$\beta$ of the central spaxel of region N. The FWHM derived for this component is 1380\,km\,s$^{-1}$, and the centroid velocity is -490\,km\,s$^{-1}$. This wing is not detected in the other emission lines.

\subsection{Dust reddening}
When calculating the dust reddening and flux ratios of NGC 1052, in order to avoid possible degeneracies introduced by the the emission-line fitting procedure, we decided to sum the fluxes of the narrow and IW components. The main reason behind this choice is the dominance of the IW component, with an integrated flux between 5 and 10 times larger if compared to the narrow component (see Fig~\ref{fig:EL}), which resulted in a low S/N in the narrow component maps.
\par
\begin{figure}
	\includegraphics[width=\columnwidth]{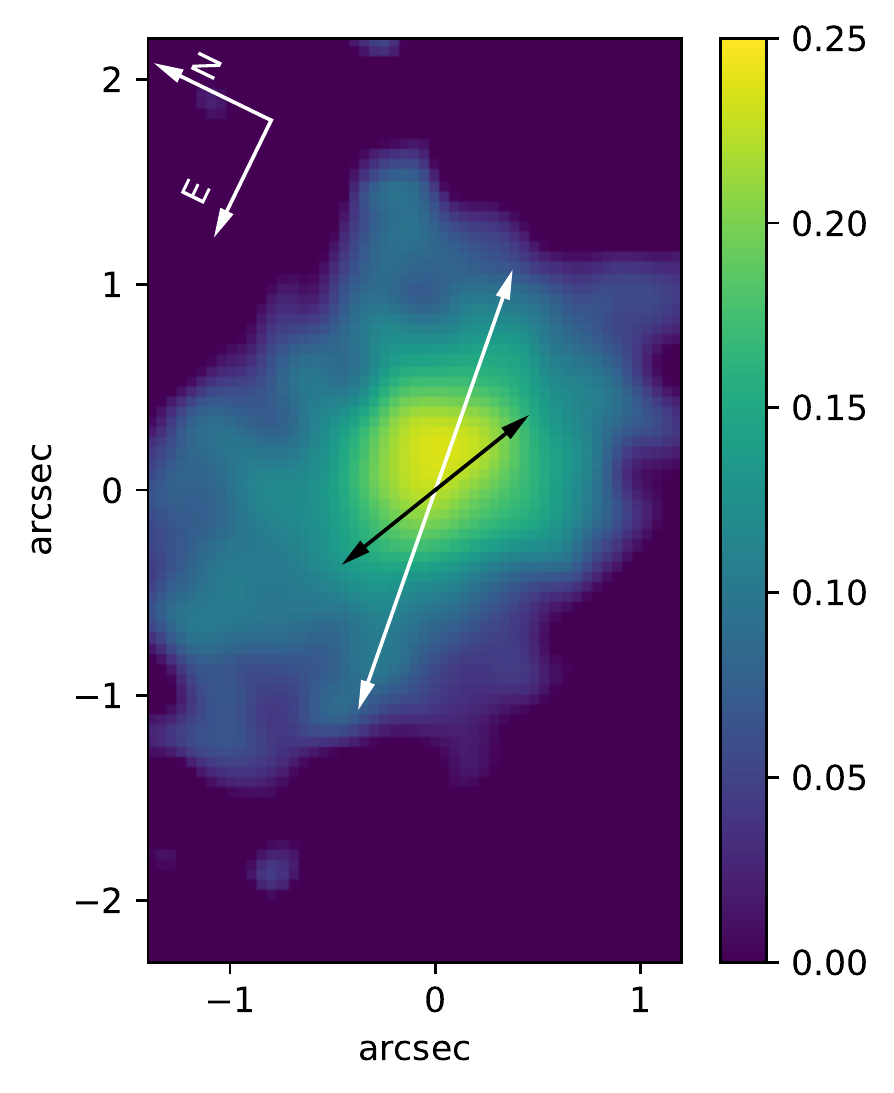}
	\caption{E(B-V) map for NGC\,1052, derived from the H$\beta$/H$\alpha$ line ratio.}
	\label{fig:ebv}
\end{figure}

In order to estimate the dust reddening [E(B-V)], we followed \citet{brum+19} and used the lines of H$\alpha$ and H$\beta$ applying the relation:

$$E(B-V) =  \frac{0.81}{ q(H\alpha) - q(H\beta)} log(\frac{H\alpha}{3.1 H\beta}) $$

with $q(H\alpha) - q(H\beta) = 0.586$ obtained from the \citet{ccm} law and assuming 3.1 as the intrinsic ratio H$\alpha$/H$\beta$ \citep[which is typical for AGNs according to][]{OsterbrockFerland06}. The map is presented in Fig.~\ref{fig:ebv}.

\subsection{Electron temperature and density}
\label{sec:temden}
In order to account for dust extinction, we corrected all optical emission lines by the E(B-V) values of  Fig.~\ref{fig:ebv}. The electron temperature was then determined based on the [O{\sc iii}]($\lambda$4959+$\lambda$5007)/$\lambda$4363  and [N{\sc ii}]($\lambda$6548+$\lambda$6583)/$\lambda$5755 ratios. To compute the electron density, we used the [S{\sc ii}]$\lambda$6716/$\lambda$6731 ratio \citep{OsterbrockFerland06}.  The calculations were performed using the {\sc temden} package present in the {\sc iraf} software \citep{ShawDufour95}.
\par
\begin{figure}
	\includegraphics[width=\columnwidth]{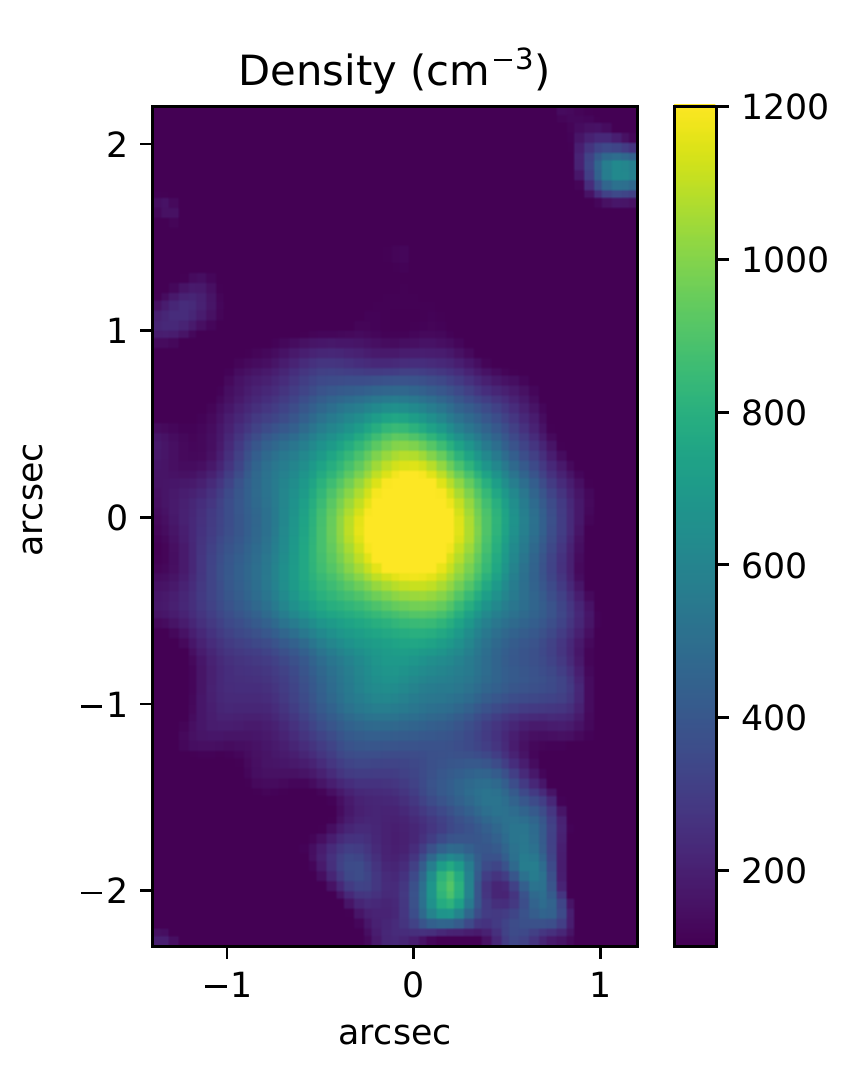}
	\caption{Density maps measured from the [S{\sc ii}]$\lambda$6716/$\lambda$6731 ratio.}
	\label{fig:den}
\end{figure}

We show the density map in Fig~\ref{fig:den}, which was estimated assuming T$_e$=10.000\,K. Since auroral emission lines are very feeble compared to their nebular counterparts, resulting in an overall low S/N, we decided to calculate temperature based in the integrated spectra of the regions N, A and B. We also extracted the integrated spectra of the circumnuclear (C) region, which we defined as the integrated datacube spectrum minus the spectrum of region N. In order to calculate these temperatures, we estimated the density based on the average map value within that region. The values of [O{\sc iii}] and [N{\sc ii}] temperatures are listed in Table~\ref{tab:temden}.
\par
As mentioned earlier, in the nuclear region, a blue wing is present in the nebular [O{\sc iii}] emission lines. We found that adding another gaussian function in the fitting of auroral [O{\sc iii}] emission produced degenerate results, between 14,000 and 30,000\,K. For that reason, we do not present [O{\sc iii}] temperature for the nuclear region.

\begin{table}
\centering
\begin{tabular}{ccc} 
 \hline
 Region & T$_{[OIII]}$ & T$_{[NII]}$\\
        &   (10$^3$K)         &    (10$^3$K)    \\
 \hline
 N      &    ---            &   12.75 $\pm$ 1.5  \\
 A      &    24.7 $\pm$ 4.3 &   10.9 $\pm$ 0.8   \\
 B      &    ---            &   ---              \\
 C      &    25.2 $\pm$ 2.8 &   11.4 $\pm$ 1.1   \\
 \hline
\end{tabular}
\caption{Temperature and density of regions N, A, B and the integrated circumnuclear (C) regions.}
\label{tab:temden}
\end{table}
%   Ha   Hb   Reddening
%N  599  127  0.34
%A  172  46   0.15
%B  80   20   0.20

%C 3336  802  0.23
\subsection{Diagnostic Diagrams}

Since we are testing the nature of the gas emission in NGC\,1052, the fact that LINER-like line ratios can be produced by old stellar populations also needs to be tested. One way of doing this is by using the WHAN diagram \citep{CF11}, defined as the equivalent width of the H$\alpha$ emission versus the [N{\sc ii}]/H$\alpha$ ratio. This diagram is divided in four regions: star forming, strong AGN, weak AGN and retired galaxies (i.e. galaxies that have stopped forming stars and are ionized by their hot low-mass evolved stars).
\par
\begin{figure*}
	\includegraphics[width=\textwidth]{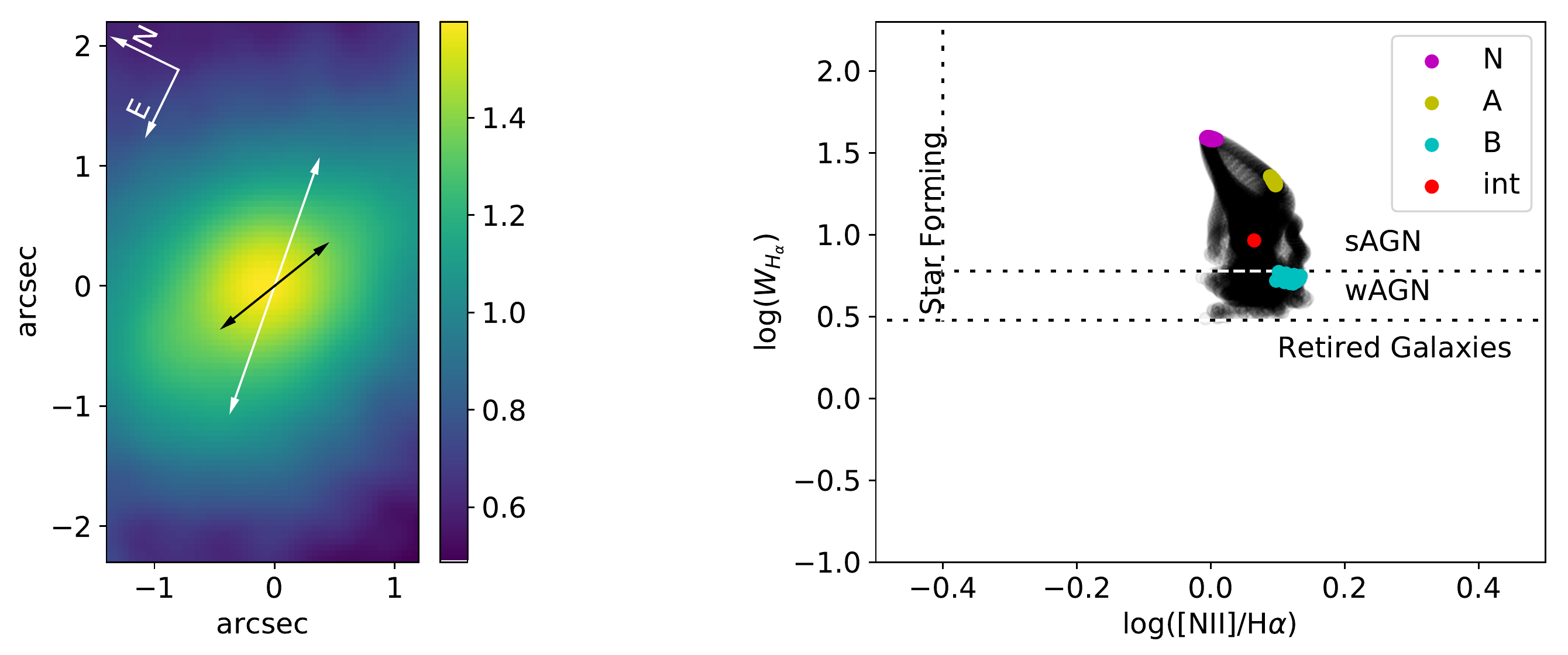}
	\caption{Left: EW$_{H\alpha}$ map. Right: WHAN diagram of NGC\,1052. Over this panel, we highlighted regions A, B and N. Also, we plotted in red the average values for the collapsed datacube.}
	\label{fig:whan}
\end{figure*}

Since the H$\alpha$ equivalent width is defined as the H$\alpha$ flux divided by the average stellar flux below its emission, we used the stellar flux derived in Paper I. Also, since both the stellar and gas emission are affected by reddening, we used both fluxes before dereddening them. WHAN diagram is presented in Fig~\ref{fig:whan}. We highlighted regions A, B and N. Also, we plotted in red the average values of the collapsed datacube.

\begin{figure}
	\includegraphics[width=\columnwidth]{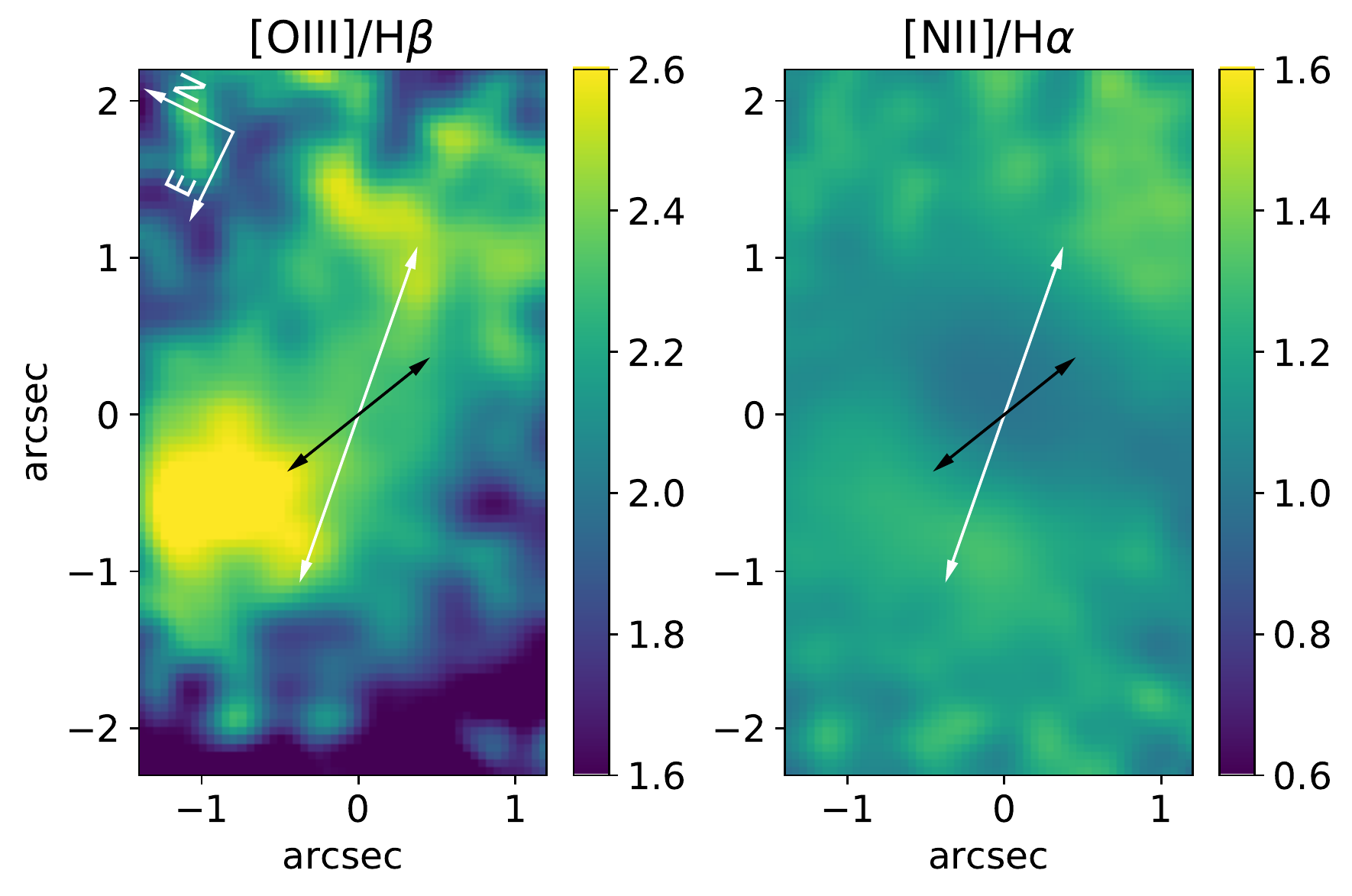}
	\caption{[O{\sc iii}]$\lambda$5007\r{A}/H$\beta$ and [N{\sc ii}]$\lambda$6587\r{A}/H$\alpha$ maps for NGC 1052. In the left panel, the white arrow indicates the orientation of the kiloparsec scale radio jet and the black arrow indicates the orientation of the parsec scale jet.}
	\label{fig:bpt}
\end{figure}

\begin{figure*}
	\includegraphics[width=\textwidth]{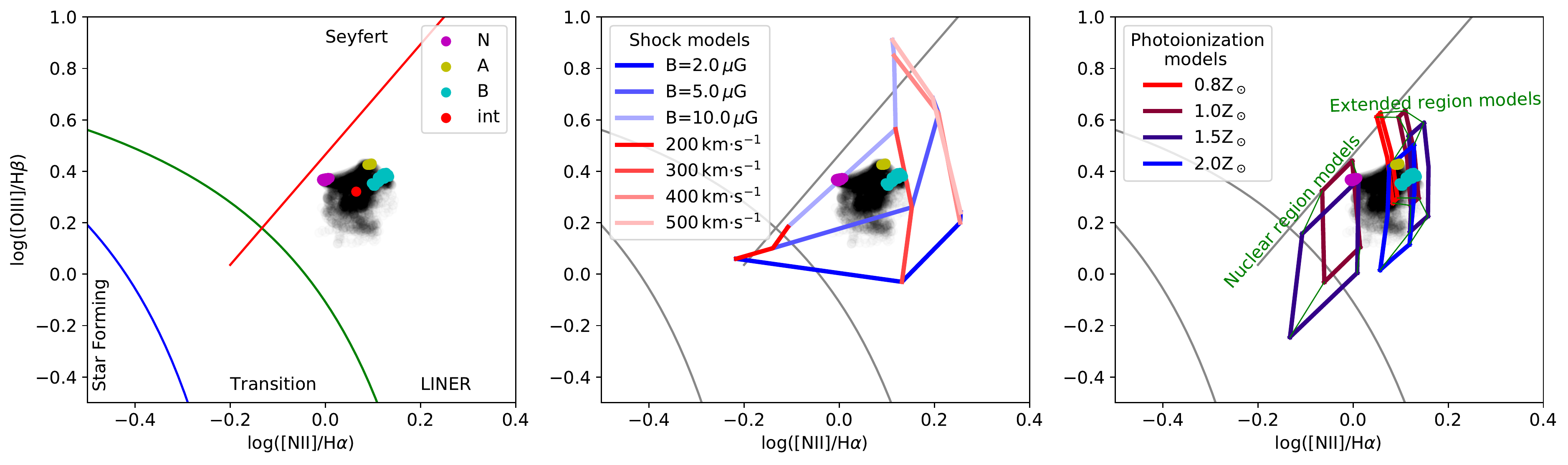}
	\caption{BPT diagram of NGC\,1052. The three main regions from Fig.~\ref{fig:EL} are highlighted (magenta, yellow and cyan), as well as the integrated average of the galaxy (red). The other regions are plotted in black. In the first panel we plotted the lines of K01 (blue), K03 (green), and K03 (red). In the middle panel, we present shock models from \citet{allen+08}. In the right panel, we plot photoionization models generated with {\sc cloudy}. Each metallicity was plotted with the same color, and green lines connect regions with the same density and ionization parameter.}
	\label{fig:diagnostics}
\end{figure*}

We also employed BPT diagrams in order to understand the nature of the emission of NGC\,1052. These diagrams uncover differences that do not appear in the flux maps. The maps of [O{\sc iii}]$\lambda$5007\r{A}/H$\beta$ and  [N{\sc ii}]$\lambda$6587\r{A}/H$\alpha$ are shown in Fig~\ref{fig:bpt}, and the respective diagnostic diagram is plotted in Fig~\ref{fig:diagnostics}. In order to disentangle the nature of NGC\,1052 emission, in the first panel we superimposed the lines from the references K01 (blue), K03 (green), and K03 (red). These lines divide the diagram in four regions based on the main ionization mechanism: Star Forming, LINER, Seyfert and transition region. The three key regions of the galaxy (A, B and N, defined in Section~\ref{sec:fitting}) are colored and distinguished from the other regions. We also collapsed the datacube and plotted the average ratios in red. In all diagnostic diagrams, the emission-line ratios from all regions of NGC\,1052 fall in the region occupied by LINERs.

\par

 To further help the characterization of NGC\,1052, over the central panel of Fig~\ref{fig:diagnostics} we superimposed sequences corresponding to \citet{allen+08} shock models with solar metallicity and preshock density of 1 cm $^{-3}$. We created our grid with models of velocities between 200 and 500\,km\,s$^{-1}$, and magnetic fields between 2.0 and 10.0\,$\mu$G.
 
 \par
Lastly, we over plotted in the right panel of Fig~\ref{fig:diagnostics}, photoionization models generated with version 17.01 of {\sc cloudy}, last described by \citet{cloudy}. In order to generate these models, we followed \citet{ricci+15} and assumed a plane-parallel geometry, a power law continuum with $f_\nu\propto\nu^{-1.5}$, which is typical for LINER-like AGNs according to \citet{ho08} and a filing factor of $10^{-3}$. We constructed two model grids, one for the nucleus and one for the extended regions. For the nucleus, we constructed our model grid with density values of 200 and 1500\,cm$^{-3}$ and a lower cut in the energy of the ionizing photons of 0.124\,eV (the default value). For the extended region, the model grid was constructed with density values of 50 and 200\,cm$^{-3}$ and a lower cut in the energy of the ionizing photons of 27\,eV (photons with less energy than 27\,eV have already been absorbed by the most internal regions of the nucleus). Each grid was constructed with ionization parameter (U) of 3.4 and 3.6 \citep[in agreement with the values derived previously for LINER-like AGNs][]{FerlandNetzer83,HalpernSteiner83,ho08} and metalicites of 0.8\,Z$_\odot$, 1.0\,Z$_\odot$, 1.5\,Z$_\odot$ and 2.0\,Z$_\odot$. For the other parameters, we used {\sc cloudy}'s default values.

\begin{figure*}
	\includegraphics[width=\textwidth]{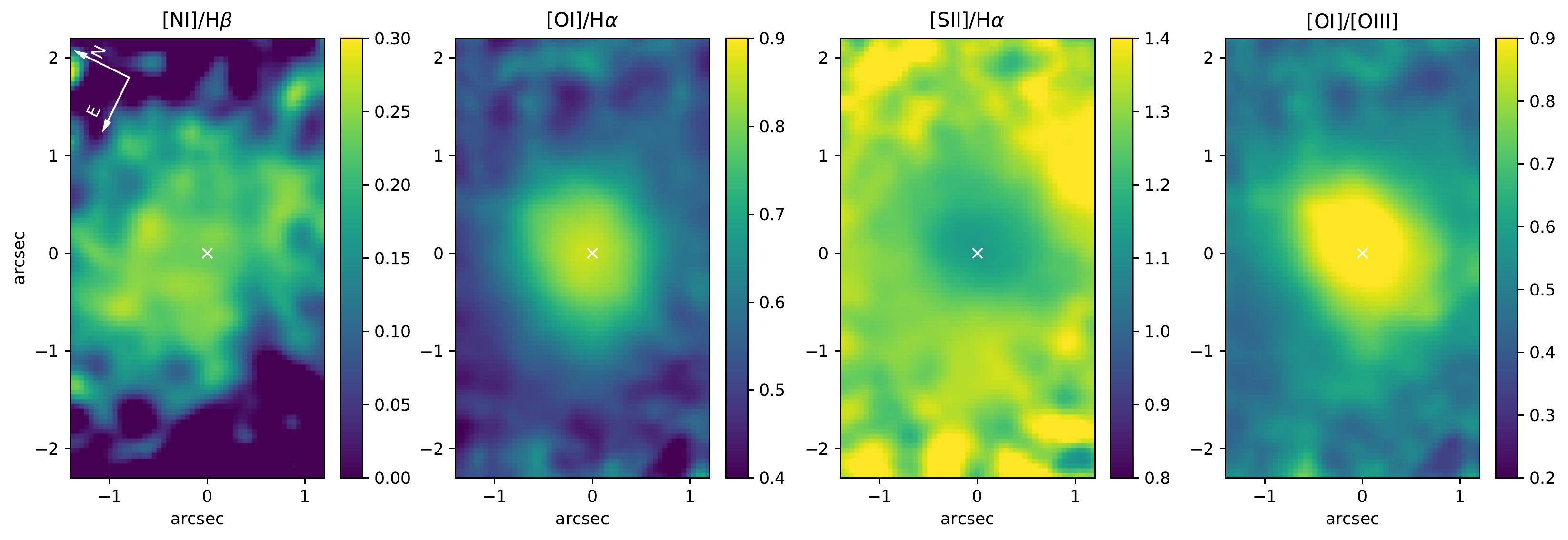}
	\caption{[N{\sc i}]/H$\beta$, [O{\sc i}]/H$\alpha$, [S{\sc ii}]/H$\alpha$ and [O{\sc i}]/[O{\sc iii}] maps corrected by dust reddening.}
	\label{fig:ratios}
\end{figure*}

In order to show how other line ratios vary over our FoV, we have also obtained emission line ratio maps of [N{\sc i}]$\lambda$5200/H$\beta$, ([O{\sc i}]$\lambda$6300+$\lambda$6360)/H$\alpha$, ([S{\sc ii}]$\lambda$6716+$\lambda$6731)/H$\alpha$ and ([O{\sc i}]$\lambda$6300+$\lambda$6360)/([O{\sc iii}]$\lambda$4959+$\lambda$5007), which are presented in Fig~\ref{fig:ratios}.

\subsection{Principal Component Analysis}

One way to analyze data cubes is by means of the PCA Tomography technique \citep{pca}. In short, this procedure uses Principal Component Analysis to search for correlations between $m$ spectral pixels across the $n$ spatial pixels of data cubes. Meaningful information are stored in a few number of eigenvectors (or eigenspectra) whose associated variances are greater than the average noise of the data cube. Tomograms are related to the projection of the eigenvectors onto the data cube, i.e. it shows where the correlations between the wavelengths occur in the spatial dimension. This technique has shown useful in extracting information from data cubes \citep{ricci+14a,menezes+13b}.
\par
\begin{figure*}
  \includegraphics[width=\textwidth]{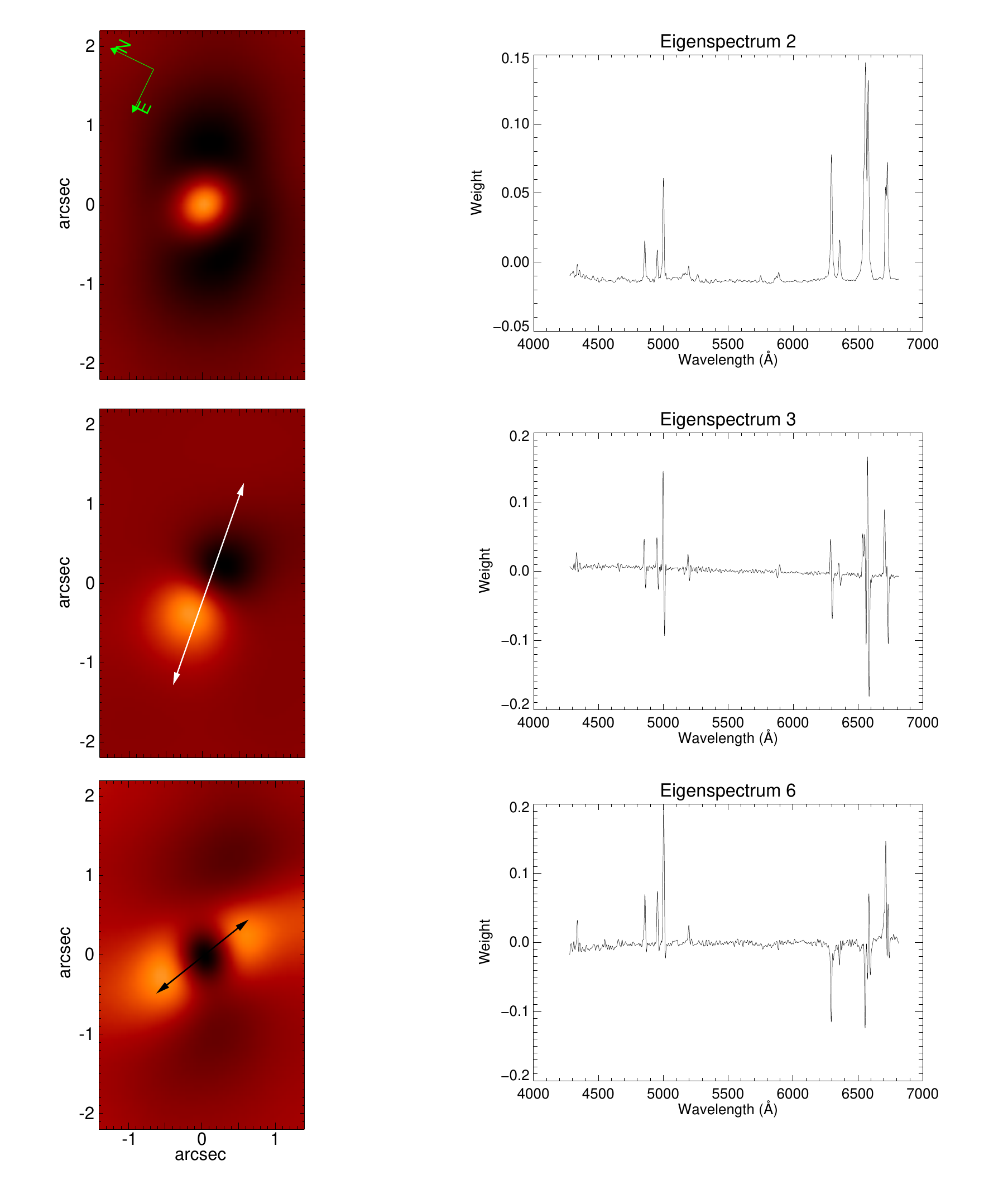}
  \caption{Eigenvectors 2, 3 and 6 for the PCA of NGC 1052. Over Eigenvector 3 map we plotted in white the orientation of the kiloparsec scale radio jet and over Eigenvector 6 map we plotted in black the orientation of the parsec scale jet.}
  \label{fig:pca}
\end{figure*}

We applied PCA Tomography to the GMOS data cube of NGC\,1052. Eigenvectors 2, 3 and 6 and their associated tomograms are presented in Fig~\ref{fig:pca}. They contain 2.4, 0.59 and 0.023\,\% of the total variance of the data cube respectively. Only these eigenvectors are shown as they are relevant to the discussion of this paper. The directions of kiloparsec and parsec scale radio jets are shown in tomograms 3 and 6, respectively.
\par On the other hand, we did not include the results from the PCA Tomography applied to the NIFS datacube, since all useful information is contained in the first eigenspectrum, which represents the bulge of the galaxy.

\subsection{Near infrared emission lines}
\begin{figure*}
  \includegraphics[width=\textwidth]{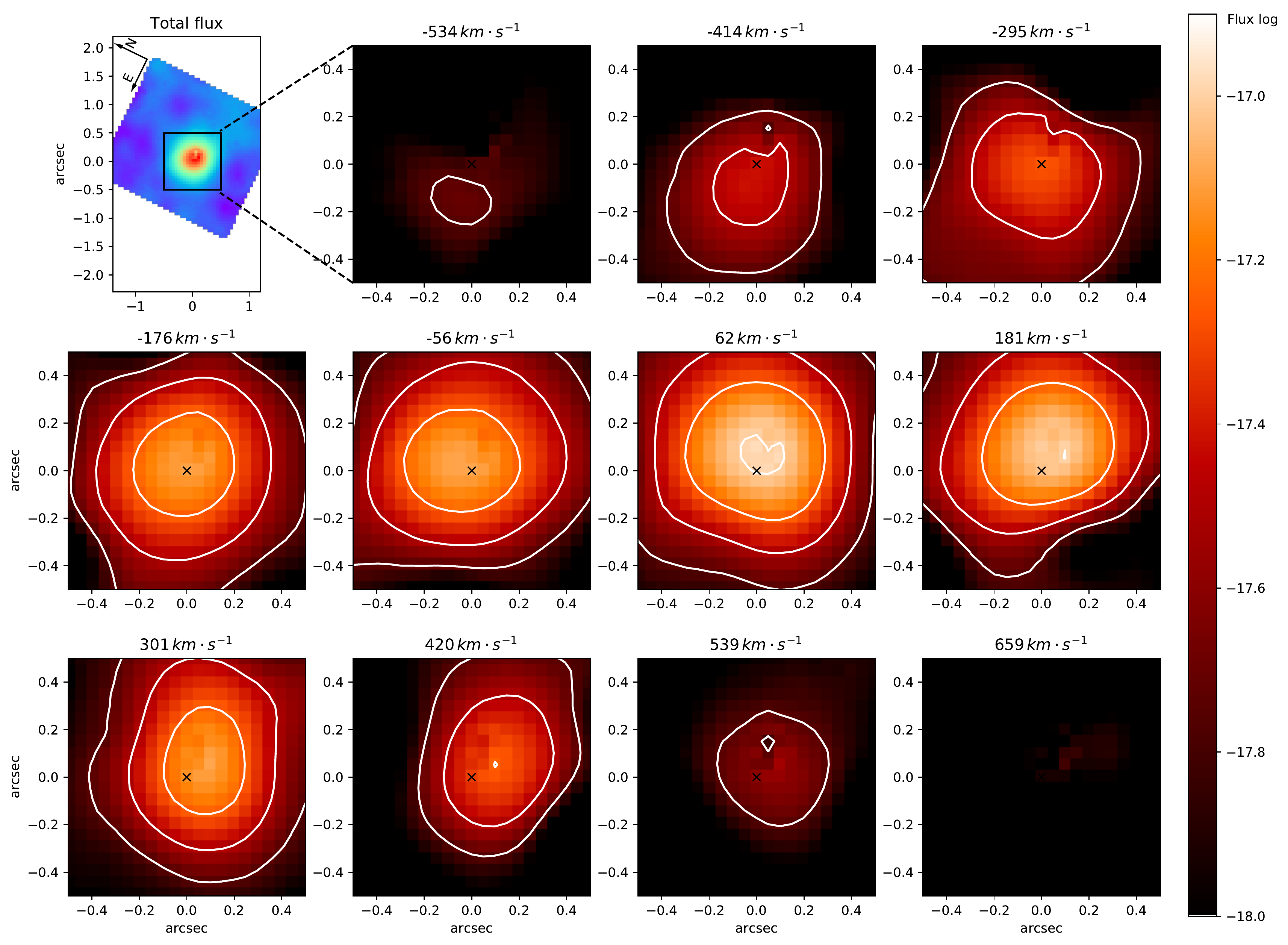}
  \caption{[Fe{\sc ii}]$\lambda$12570\r{A} emission line flux and channel maps.}
  \label{fig:FeII}
\end{figure*}

\begin{figure*}
  \includegraphics[width=\textwidth]{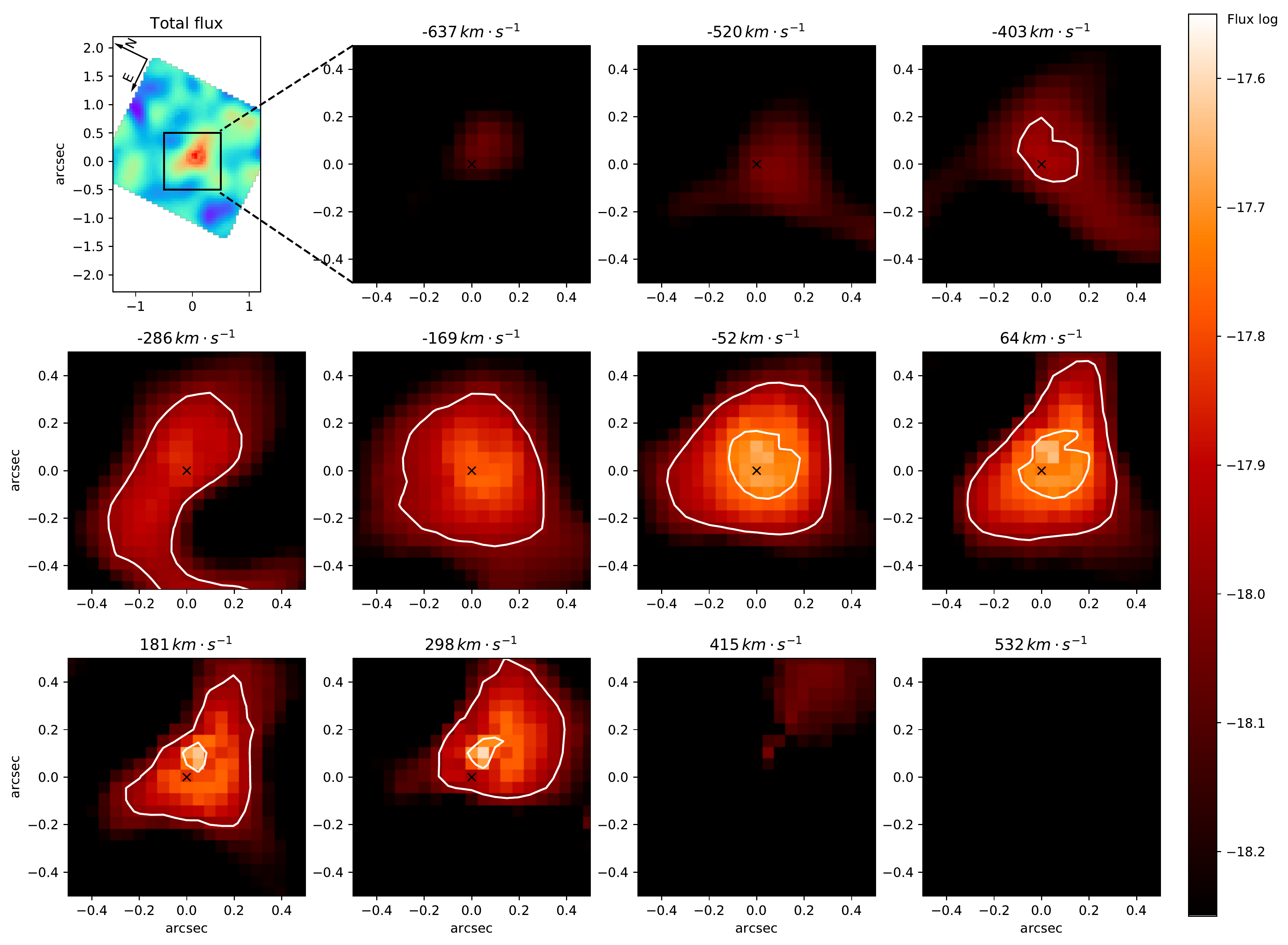}
  \caption{Pa$\beta$ emission line flux and channel maps.}
  \label{fig:PaB}
\end{figure*}

\begin{figure*}
  \includegraphics[width=\textwidth]{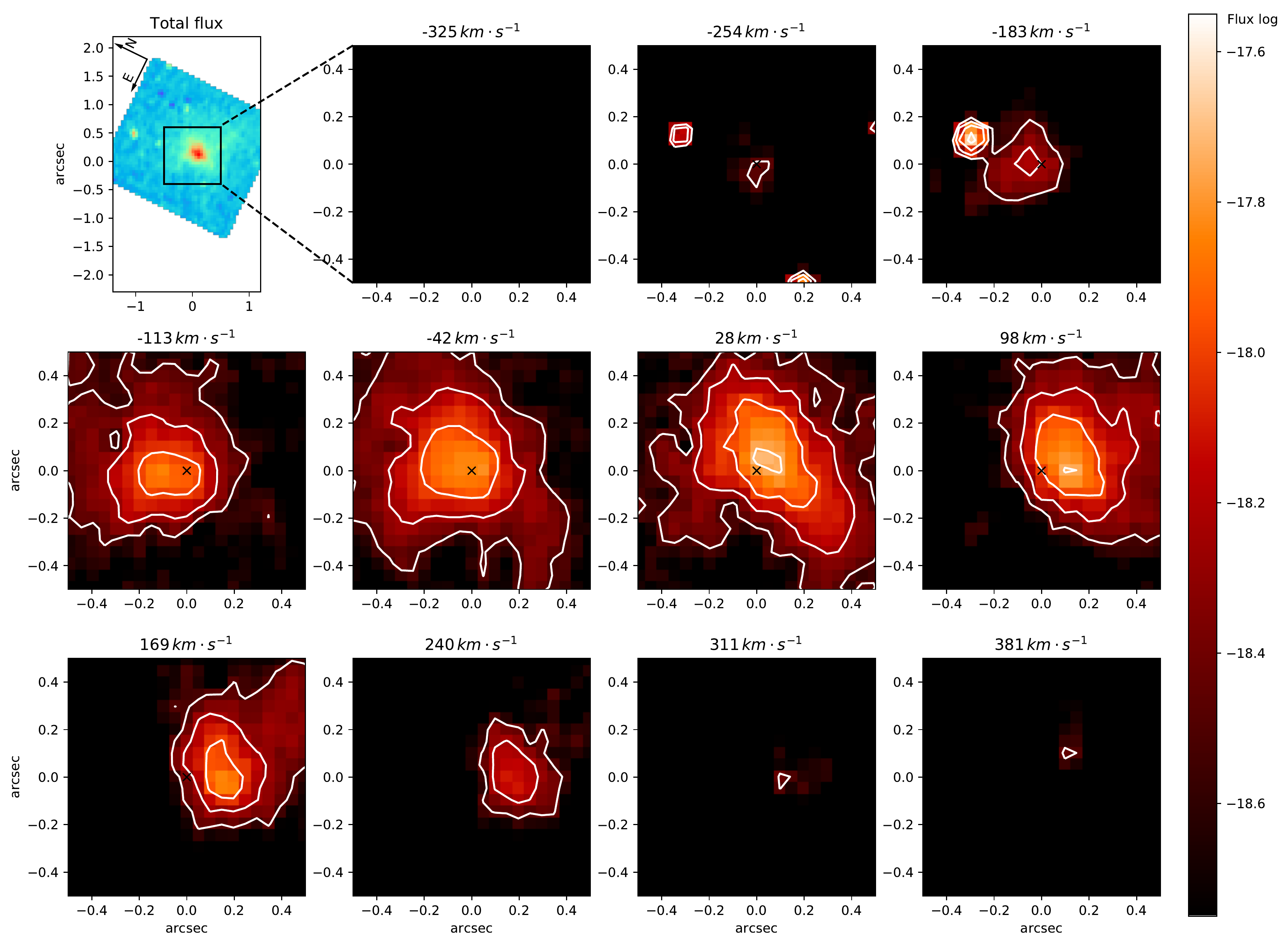}
  \caption{H$_2$ $\lambda$2121\r{A} emission line flux and channel maps.}
  \label{fig:H2}
\end{figure*}

\begin{figure}
  \includegraphics[width=\columnwidth]{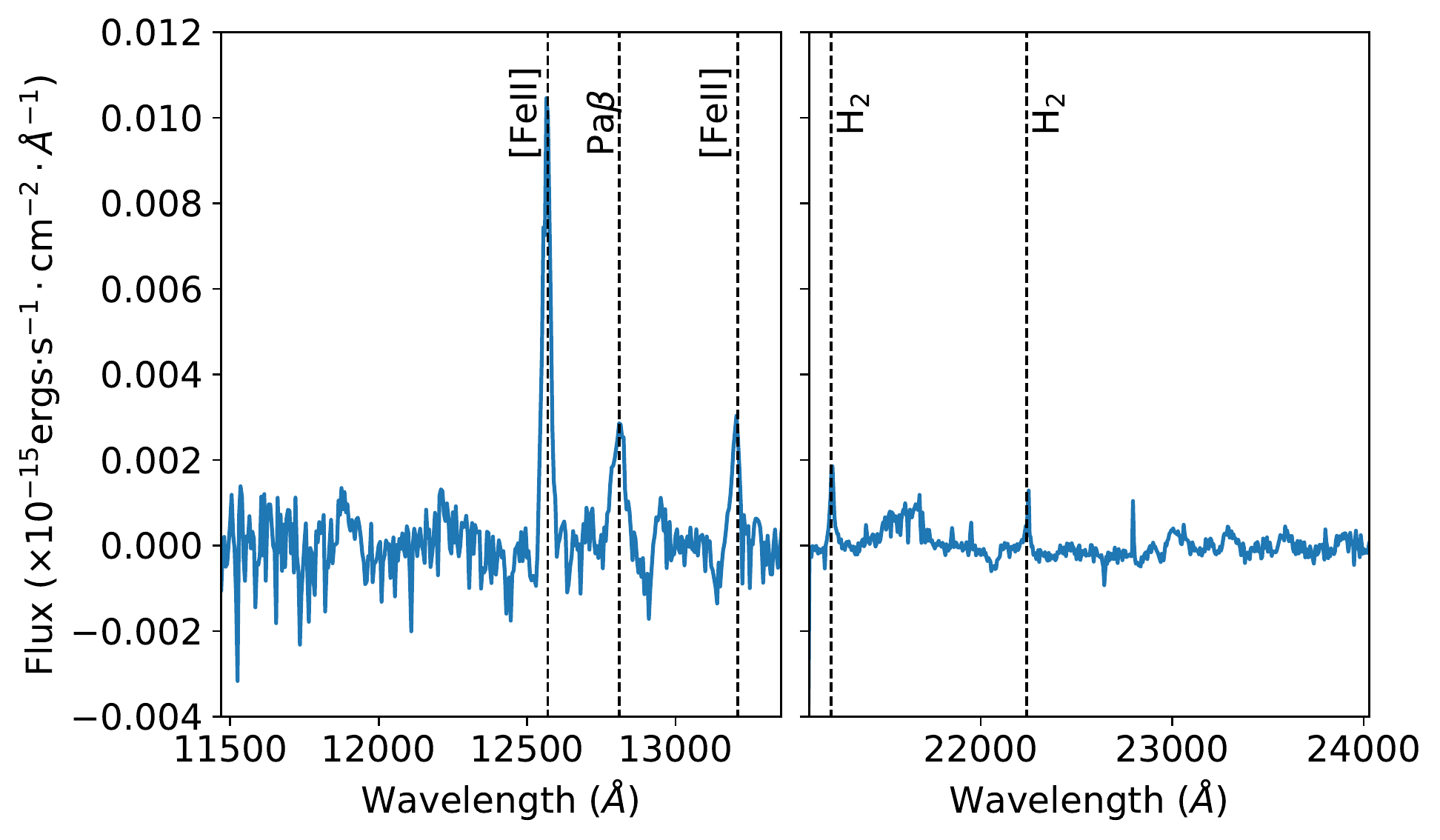}
  \caption{J and K band NIR spectra of the central spaxel of region N. The main emission lines are indicated.}
  \label{fig:nirspec}
\end{figure}

Five emission lines were detected in the NIR datacubes: [Fe{\sc ii}]$\lambda$12570\r{A}, Pa$\beta$, [Fe{\sc ii}]$\lambda$13210\r{A}, H$_2$ $\lambda$21218\r{A} and H$_2$ $\lambda$22230\r{A}. We present in Fig.~\ref{fig:nirspec}, the J and K band spectra of the central spaxel of region N. Since the [Fe{\sc ii}]$\lambda$13210\r{A} and H$_2$ $\lambda$22230\r{A} lines are weaker than the $\lambda$12570\r{A} and $\lambda$21218\r{A} lines, and both probe the same gas portion, we only show the maps of the later ones. Given the compactness of NIR emission lines, we decided to analyze this data through channel maps. This alternative method was required, since direct emission line fitting did not reveal any structure in either centroid velocity or FWHM maps. The emission line fluxes and channel maps of [Fe{\sc ii}]$\lambda$12570\r{A}, Pa$\beta$ and H$_2$ $\lambda$21218\r{A} are presented in Figs~\ref{fig:FeII} to \ref{fig:H2}. We chose a constant bin size for the channels of 5\r{A}, because it is a multiple of the binning after the synthesis, also providing a good S/N. 

\begin{figure*}
  \includegraphics[width=\textwidth]{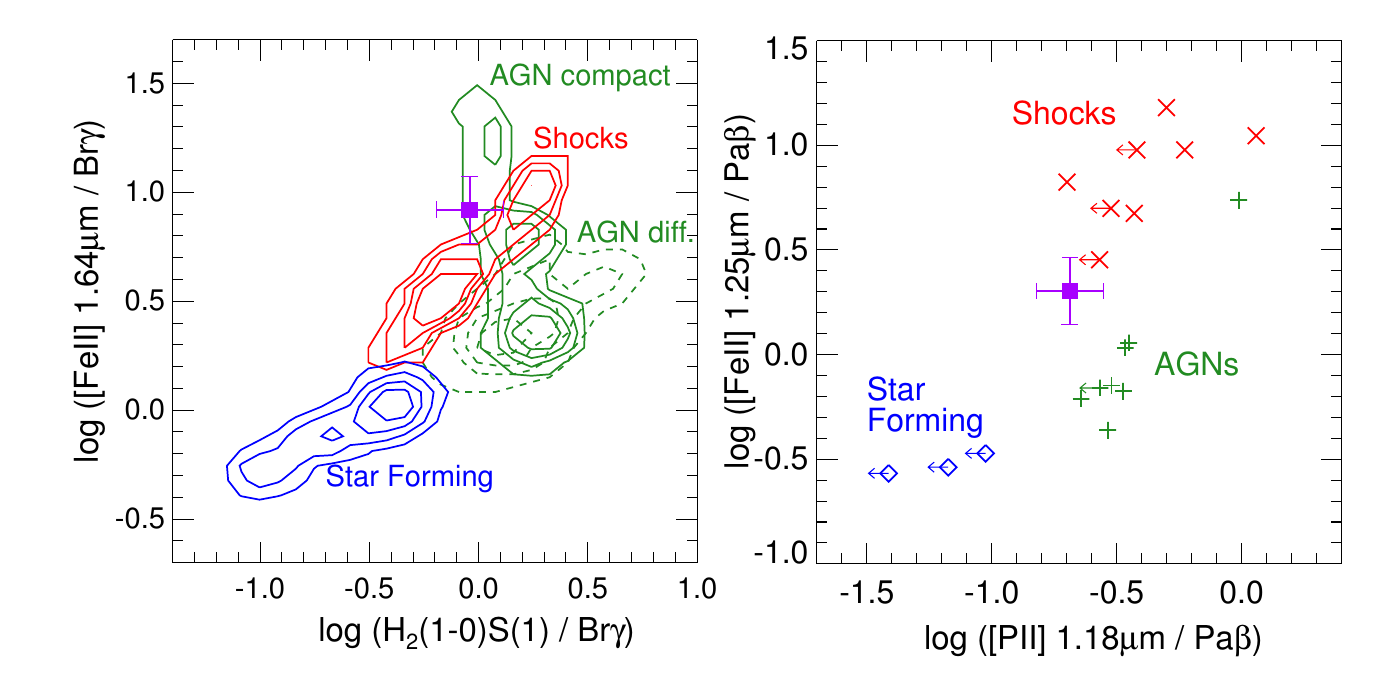}
  \caption{NIR emission line ratio diagrams from \citet{maiolino+17}. In both diagrams, the position of NGC\,1052 is plotted in purple with error bars.}
  \label{fig:NIRdiagrams}
\end{figure*}

The NIR emission-line ratios can also be used as diagnostic diagrams to discriminate the dominant ionization mechanism of emission line galaxies \citep[see for example][]{larkin+98,rodriguez-ardila+05,riffel+13,colina+15}. In order to improve the determination of the dominant ionization mechanism behind NGC\,1052, we present in Fig~\ref{fig:NIRdiagrams} two emission-line ratio diagrams from \citet{maiolino+17}. The purple square shows NGC\,1052 position in these diagrams. Since a few of these lines are not in the wavelength range of our datacube, we measured the line fluxes from the NIR integrated spectrum of NGC\,1052 presented by \citet{mason+15}. We performed these measurements after subtracting the stellar content of the spectrum, which was done by applying the same method described in Section~\ref{sec:fitting} (i.e. E-MILES models and {\sc starlight} code).

\section{Discussion}
\label{sec:discussion}
\subsection{Fluxes and Kinematics}
\par
From Fig~\ref{fig:EL}, it is possible to see that narrow components have a flux distribution which is oriented in the northeast/southwest direction. The kinematics of these emission lines are also aligned in the same direction, dividing the galaxy into northeast (blueshifted) and southwest (redshifted). A similar result was already reported by D15. Since this component extends beyond the edges of our FoV, with the same orientation, our narrow components are probably the same components observed by D15. According to their analysis, the best explanation for this data is the presence of two gas bubbles.
\par
The [O{\sc iii}] fluxes are also oriented in the northeast/southwest direction. However, the flux peak of [O{\sc iii}] is off-centered $\sim$0\farcs35 to the west of the continuum peak, in the direction of the parsec scale jet. Also, the kinematics of the [O{\sc iii}] emissions are compatible with the other ones, with the exception of one region. To the west of our FoV, following the direction of the kiloparsec scale radio jet, the [O{\sc iii}] narrow kinematics grows to higher values if compared to the other narrow emissions. These results seem to indicate that to the west of our FoV, narrow [O{\sc iii}] is probably tracing the interaction of the kiloparsec radio jet with the environment.
\par
The fluxes of IW components, in contrast, present approximately elliptical isocontours with a major axis oriented between kiloparsec and parsec scale radio jet directions. In addition, the kinematics of all IW components are similar to each other, with compatible FWHM and centroid velocity distributions. From Fig~\ref{fig:EL}, it is possible to see that the blueshift velocities of this component reach their maximum in region A, beginning to fall afterwards. The redshifted portion, on the other hand, reaches its maximum in the borders of our FoV, suggesting that it might keep increasing. However, since it was not detected by the large (40\farcs0$\times$25\farcs0) FoV of D15, it suggests that this component does not extend much further away from our FoV. 
\par
Also, in the [O{\sc iii}] FWHM map, close to  (0\farcs, -1\farcs), a drop in the FWHM is found, co-spatial with a rise in the velocity field of the [O{\sc iii}] IW component. This happens because the fitting in this region is highly degenerate, as can be seen in Fig~\ref{fig:fits}. However, there is no clear sign of a difference in the emission line behaviour in this same region.
\par
One possible explanation for the kinematics of the IW is that it is caused by a bipolar outflow. In this scenario, the narrow components would trace the gas bubble identified by D15, whereas the IW component would trace an on-going outflow related to the radio jets. Supporting this hypothesis is the fact that this component is oriented along a direction between those of the parsec and kiloparsec radio jets.
\par
\begin{figure*}
  \includegraphics[width=\textwidth]{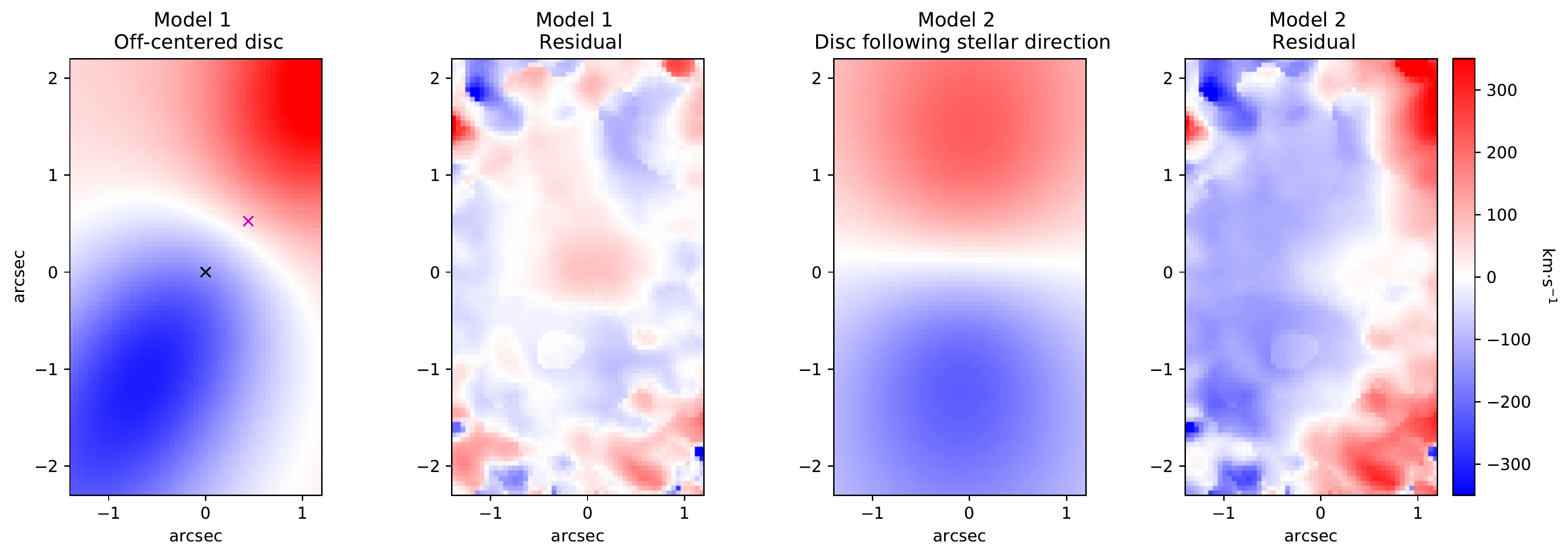}
  \caption{Two disc models of the H$\beta$ IW kinematics with their respective residuals. The first model was obtained with all parameters free to vary, whereas the second model was obtained by setting the orientation of the nodes line, disc inclination and conematic center to the ones modeled for the stars. The black x in the second panel indicates the luminosity peak, whereas the magenta x indicates the center of the rotation.}
  \label{fig:IW}
\end{figure*}

In the above scenario, no component is tracing the gravitational potential of the galaxy. Thus, in order to test if this component can also be explained by a disc in circular motion, we followed \citet{bertola+91} and assumed that the gas follows circular orbits in a plane with a rotation curve given by :

\begin{equation} 
v_c = \frac{Ar}{(r^2 + c_0^2)^{p/2}},
\end{equation} 
\noindent
where $r$ is the radius and $A$, $c_0$ and $p$ are parameters to be determined. The observed radial velocity at a position ($R$,$\Psi$) on the plane of the sky is then given by:

\scriptsize
\begin{equation}
v(R, \Psi) = v_{sys} + \frac{AR cos(\Psi-\Psi_0)sin\theta cos^p \theta}{\{R^2[sin^2(\Psi-\Psi_0)+cos^2\Theta cos^2(\Psi-\Psi_0)]+c_0^2 cos^2\Theta  \}^{p/2}},
\end{equation}
\normalsize
\noindent
where $\Theta$ is the disc inclination ($\Theta$=0 being is a face-on disc), $\Psi_0$ is the position angle of the line of nodes and $v_{sys}$ is the systemic velocity of the galaxy. We have fitted the above expression to the H$\beta$ IW velocity field using our own script, which automatically searches for the center, inclination and velocity amplitude and has already been used in Paper I. The resulting model and the respective residuals are presented in the first two panels of Fig~\ref{fig:IW}.

The fact that the model fits properly the entire FoV of the IW component, with all residuals smaller than 50\,km\,s$^{-1}$, indicates that the kinematics of the IW component is well explained by a disc following circular motion. However, the value of two parameters indicate that, if this is indeed the case, the disc is not circular around the SMBH. First, as can be seen in Fig~\ref{fig:IW}, the center of the rotation (magenta x) is located 0\farcs7 west of the continuum  luminosity peak (black x), adopted as the location of the galaxy nucleus. Second, the systemic speed of this galaxy, as derived from the fitting of the IW kinematics, is 68\,km\,s$^{-1}$ higher than the systemic speed derived from the stellar model. Such eccentric discs have already been reported in the literature for other galaxies \citep[e.g. M31][]{menezes+13a}. If that is the case, it is probably a consequence of the merger event experienced $\sim$1\,Gyr ago with a gas-rich dwarf or spiral \citet{gorkom+86}. 
\par
A third option to explain the IW gas kinematics is the combination of a rotation and an outflow component. In order to test this possibility we performed the fit of the data again, but setting the orientation of the line of nodes, the disc inclination and the kinematic center to those obtained from the fit of the stellar velocity field (Paper I). The resulting model and the respective residuals are presented in the last two panels of Fig~\ref{fig:IW}. Looking at the residuals, it is possible to see that they are similar to the kinematics of the narrow components, but with higher values. One possible explanation for these results is that the blue residuals originate in the front part of the blue bubble, whereas the narrow kinematics would originate in the rear part of this bubble. Also, the red residuals would originated in the rear part of the red bubble, whereas the narrow kinematics would originate in the front part of this bubble. In this scenario, the radio jets would pass inside the bubbles with the SMBH and its torus located in between these bubbles.
\par
Combined, these results allow us to conclude that NGC\,1052 has at least five different kinematic components: i) the unresolved broad component visible in H$\alpha$, limited to the nucleus, originating in the BLR; ii) the unresolved intermediate-broad [O{\sc iii}], also limited to the nuclear region; iii) the IW component, detected in all emission lines and which is tracing either an outflow, an eccentric disc or a combination of a disc with large scale shocks; iv) the narrow components of the H{\sc i}, [O{\sc i}], [N{\sc i}], [N{\sc ii}] and [S{\sc ii}], tracing the gas bubbles first detected by D15; v) the [O{\sc iii}] narrow component, which follows the fluxes and kinematics of the other narrow emission lines everywhere but to the west, where it is probably tracing the interaction of the kiloparsec radio jet with the environment \par
The presence of three distinct kinematic components in the emission line profiles of the nuclear spectrum of NGC\,1052 has already been reported by S05, who analyzed the inner 3\farcs0x3\farcs0 region of the galaxy. Our results not only confirm their hypothesis, but also show that the kinematics are even more complex, with an unresolved [O{\sc iii}] component in the nucleus and a narrow [O{\sc iii}] component detached from the other narrow components to the west.
\par
FWHM maps, on the other hand, do not highlight any major differences, mainly due to the narrow range we allowed them to vary. This narrow range was needed, since allowing higher FWHM variations caused {\sc ifscube} to not fit properly the noisy regions of the datacube. 

\subsection{Dust Reddening}

The E(B-V) map peaks at $\sim$0.25\,mag inside the nuclear region N (see Fig~\ref{fig:ebv}). This value is lower than the values derived by both \citet[][E(B-V)=0.44]{gabel+00} and D15 (E(B-V)=0.33). However, this difference is not significant, since the uncertainties in E(B-V) are high due to the difficulty in separating H$\alpha$ from the adjacent [N{\sc ii}] lines.
\par
In the jets direction, the reddening is lower than in the vicinity of the nucleus, with an average value of 0.1\,mag, but higher than the rest of the FoV, where the E(B-V) is negligible. These results show that this galaxy is not very affected by dust reddening, which is consistent with past results showing that this is the case for most elliptical galaxies \citep{PadillaStrauss08,zhu+10}.
\par
%In paper I, we derived a dust extinction of A$_V$=0.5\,mag for the stars in region N. Assuming A$_V$=3.1\,E(B-V), it translates to a stellar E(B-V) of 0.16\,mag. The difference in the reddening of the gas and the stars is small, and might be caused by noise alone.

\subsection{Electron Temperature and Density}
The density values  derived by us (Fig~\ref{fig:den}) reveal that the highest values are seen inside the nuclear region N. While we found a density of $\sim$1200\,cm$^{-3}$ in the region N, the surrounding regions have densities close to or below the [S{\sc ii}]/[S{\sc ii}] sensitivity limit ($\lesssim$100\,cm$^{-3}$).
\par
The temperatures detected by us (Table~\ref{tab:temden}), on the other hand, do not highlight a major difference between the nuclear region N and the surrounding regions. Temperature values measured from [N{\sc ii}] emission lines are between 10,000 and 13,000\,K, with no significant difference in region N. Temperature measured from [O{\sc iii}] have values closer to 25,000\,K in regions A and C, and could not be measured in regions N and B.
\par
The high temperatures measured from [O{\sc iii}] are too high to be compatible with photoionization \citep{OsterbrockFerland06}. However, as indicated in Section~\ref{sec:temden}, this value could not be measured in the nuclear region N. Temperatures measured from [N{\sc ii}], on the other hand, are compatible with photoionization in the three regions that we have been able to measure them. 
\par
The simple presence of a broad line profile in the H{\sc i} lines indicates that direct photoionization by the AGN must be present in the nuclear region. This is compatible with the low temperatures measured from [N{\sc ii}]. However, a contribution coming from shocks cannot be fully discarded in this region, since the fitting of the profile of the [O{\sc iii}]$\lambda$4363 emission line allows us to calculate values which are not compatible with photoionization.
\par
Outside of region N, the high [O{\sc iii}] temperatures and the low [N{\sc ii}] temperatures suggest that a single mechanism is not capable to fully explain the extended ionization observed. It is likely that both shocks and photoionization play a role in ionizing the gas in these extended regions.

\subsection{Diagnostic Diagrams}
The WHAN diagram of Fig~\ref{fig:whan} shows that the emission lines of NGC\,1052 are typical of AGNs. This is important because it rules out ionization by hot low-mass evolved stars and also young stars, thus restraining the ionization mechanism of the galaxy to be either shocks or photoionization by the AGN.
\par
Also, the [O{\sc iii}]/H$\beta$ and [N{\sc ii}]/H$\alpha$ ratios (Fig~\ref{fig:diagnostics}) confirm that the LINER-like emission extends throughout the whole FoV, not being restricted to the nucleus of the galaxy. The regions A, B and N are compatible with ionization by shocks with velocities between 200 and 300\,km\,s$^{-1}$, and a magnetic field between 5.0 and 10.0 $\mu$G (panel b of Fig~\ref{fig:diagnostics}). Besides that, region N is compatible with our photoionization models for the nucleus, whereas regions A and B are also compatible with our photoionization models for the extended region (panel c of Fig~\ref{fig:diagnostics}).
\par
From Figs~\ref{fig:bpt} and \ref{fig:ratios}, it is possible to see that region N behaves differently when compared to the rest of the galaxy. Also from these figures, it is possible to see that all ratios are constant outside the nuclear region, with the exception of [O{\sc iii}]/H$\beta$, which is slightly higher at regions A and B, if compared to the rest of our FoV. These results are compatible with our hypothesis of photoionization as the main ionizer inside region N, with the extended region of this galaxy being ionized by a combination of shocks and photoionization.

\subsection{Principal Component Analysis}

Eigenvectors that results from the PCA Tomography technique reveal the correlations between the wavelengths of the spectral dimension of a data cube. In the case of the optical data cube of NGC\,1052, the second eigenvector shows strongly the characteristics of the AGN; but also that the emission lines are correlated with the Fe\,I$\lambda$5270 and Na\,D$\lambda$5893 stellar absorption features. This correlation is located in the nucleus of the galaxy. An interpretation for this result is that the featureless continuum from the AGN is decreasing the equivalent widths of the stellar absorption lines, which was already discussed in Paper I. Similar cases were observed in the nuclei of the galaxies IC\,1459 \citep{ricci+14a,ricci+15} and NGC\,1566 \citep{dasilva+17}. In Eigenvector 3, one may see an anti-correlation between the red and the blue wings of the emission lines. This is associated with a bipolar structure seen in the spatial dimension with a position angle PA=81$\pm$1 degrees. This is usually a signature of a gas disc \citep{ricci+14a,dasilva+17}. In Eigenvector 6, emission lines from elements with a low ionization potential ([O{\sc i}] and [N{\sc ii}]) are anti-correlated with the emission lines from elements with higher ionization potential and with the H{\sc i} lines. The tomogram shows that the nucleus is dominated by the low ionization emission while a bipolar structure, associated with the [O{\sc i}] and the H I lines, is seen with a P.A. of 53$\pm$1 degree. This P.A. is similar to the direction of the parsec scale jet of NGC\,1052. Given that \citet{sawada-satoh+16} detected a molecular torus perpendicular to this jet, we propose that Tomogram 6 is related to an ionization cone, i.e. ionizing photons are collimated by the torus in the direction of the bipolar structure that is seen in this image. Another result that supports this interpretation is the fact that the [O{\sc iii}]/H$\beta$ ratio is higher in the position of the structure that is located northeast from the nucleus (Fig~\ref{fig:bpt}).
\par
It is clear that the position angle of the bipolar structure we associate with the ionization cone (PA=53$\pm$1) is different from that of the parsec scale radio jet \citep[PA$\sim$66, as  estimated by us based on the mean PA of the jet components identified by][]{lister+19}. Since the radio jet presumably originates in the inner accretion disc and the ionization cone is collimated by the molecular/dusty torus, we conclude that their projection on the sky is misaligned by $\sim$13 deg.

\subsection{NIR emission lines}
From Fig~\ref{fig:FeII}, it is possible to see that [Fe{\sc ii}] channel maps are redshifted to the east and blueshifted to the west of the nucleus. Similar to the intermediate components of the optical emission lines, it shows that the orientation of [Fe{\sc ii}] kinematics also lies between the orientation of kiloparsec and parsec scale radio jets. However, [Fe{\sc ii}] emission is much more limited to the central region, undetected beyond 0\farcs5.
\par
Pa$\beta$ emission (Fig~\ref{fig:PaB}), on the other hand, does not show any kinematic features in the channel maps, with all velocity channels showing similar flux distributions, concentrated in the central regions. Lastly, H$_2$ channel maps (Fig~\ref{fig:H2}) show that this emission line shows mostly blueshifts to the north of the nucleus and redshifts to the south, with an apparent line of nodes similar to that implied by the narrow-line kinematics. This orientation is also similar to that of the D15 gas bubbles.
\par
Fig~\ref{fig:NIRdiagrams} reveals that the NIR emission lines of NGC\,1052 are compatible with both photoionization by and AGN as well as shocks (produced by the radio jets), as seen in the two diagnostic diagrams. These results also help confirming that only one ionization mechanism is not capable of fully explaining the emission lines of NGC\,1052.

\section{Conclusions}
\label{sec:conclusions}

In this work, we studied the gas excitation and kinematics of the inner 320$\times$535\,pc$^2$ of NGC\,1052, both in the optical and in the NIR. Our results can be summarized as follows:
\begin{itemize}
\item At least five kinematic components are present in this galaxy:
\begin{enumerate}
\item In the nucleus, an unresolved broad (FWHM$\sim$3200\,km\,s$^{-1}$) line is visible in H$\alpha$.
\item Also in the nucleus, an unresolved intermediate-broad component (FWHM$\sim$1380\,km\,s$^{-1}$) is seen in the [O{\sc iii}]$\lambda \lambda$4959,5007 doublet. 
\item An IW component (280<FWHM<450\,km\,s$^{-1}$) with central peak velocities up to 400\,km\,s$^{-1}$. Possible explanations for this feature are: an outflow, an eccentric disc or a combination of a disc with large scale shocks.
\item A narrow (FWHM<150\,km\,s$^{-1}$) component, visible in the H{\sc i}, [O{\sc i}], [N{\sc i}], [N{\sc ii}] and [S{\sc ii}] emission lines, which extends beyond the FoV of our data. This component is compatible with two previously detected gas bubbles, which were attributed to large-scale shocks.
\item Another narrow (FWHM<150\,km\,s$^{-1}$) emission, visible only in [O{\sc iii}], which differs from the other narrow emission lines along the kiloparsec radio jet, probably tracing the interaction of the kiloparsec radio jet with the environment.
\end{enumerate}
\item When analyzing density, temperature and diagnostic diagrams, our results suggest that the ionization of the FoV of our data cannot be explained by one mechanism alone. Rather, our results suggest that photoionization is the dominant mechanism in the nucleus, with the extended regions being ionized by a combination of shocks and photoionization.
\item We found that the dusty molecular torus is misaligned with the inner accretion disc by $\sim$13\,deg in the plane of the sky. 
\item We confirmed the presence of an unresolved featureless continuum associated with the LLAGN
\item From NIR data, we found that [Fe{\sc ii}] is also oriented in a direction compatible with the radio jets, whereas H$_2$ is oriented in the direction of the narrow components. 
\end{itemize}

\section*{Acknowledgements}

We thank the anonymous referee for reading the paper carefully and providing thoughtful comments that helped improving the quality of the paper. LGDH, TVR and NZD thank CNPq.  RR thanks CNPq, FAPERGS and CAPES for partial financial support to this project. This research has made use of data from the MOJAVE database that is maintained by the MOJAVE team \citep{lister+18}. This study was financed in part by the Coordena\c{c}\~ao de Aperfei\c{c}oamento de Pessoal de N\'ivel Superior - Brasil (CAPES) - Finance Code 001.

\bibliographystyle{mnras}
\bibliography{luisgdh} % if your bibtex file is called example.bib

\bsp	% typesetting comment
\label{lastpage}

\end{document}